\documentclass[12pt]{article}

\newcommand{\beq}[1]{\begin{equation} \label{#1} }
\newcommand{\eeq}   {\end{equation}}
\newcommand{\ds}{\displaystyle \mathstrut}
\newcommand{\Frac}[2]{\frac{\textstyle \mathstrut #1}
{\textstyle \mathstrut #2}}
\newcommand{\kp}{\mbox{\ae}}
\newcommand{\kps}{\mbox{\small\ae}}
\newcommand{\vkappa}{\mbox{\boldmath $\kappa$}}
\newcommand{\vrho}{\mbox{\boldmath $\rho$}}
\newcommand{\av}[1]{\langle #1 \rangle}

\begin{document}

\title{Multiple scattering theory for slow neutrons (from thermal
to ultracold)}
\author{A.L.Barabanov, S.T.Belyaev\\
{\it Kurchatov Institute, 123182 Moscow, Russia}}
\date{}
\maketitle

\abstract{The general theory of neutron scattering is presented,
valid for the whole domain of slow neutrons from thermal to
ultracold. Particular attention is given to multiple scattering
which is the dominant process for ultracold neutrons (UCN). For
thermal and cold neutrons, when the multiple scattering in the
target can be neglected, the cross section is reduced to the known
value. A new expression for inelastic scattering cross section for
UCN is proposed. Dynamical processes in the target are taken into
account and their influence on inelastic scattering of UCN is
analyzed.}
\bigskip

diffraction and scattering
\bigskip

\section{Introduction}
\label{intro}
Thermal and cold neutrons with wave length
$0.03$~nm~$\leq\lambda\leq1$~nm\enskip is an important tool for
investigation of condensed matter. Due to the absence of charge
and considerably weak interaction with electrons and nuclei
incident neutron wave goes deep into target almost without
distortion and coherently influences on all atoms of the target.
The whole specific features of the matter (crystalline and
magnetic structure etc.) show themselves in interference of the
secondary scattered waves.

Theory of neutron-substance interaction for thermal and cold
neutrons is well established (see, e.g.,
\cite{Tur65,Gur68,Lov84}). It is based on the use of Fermi
pseudopotential
\beq{1.1}
V({\bf r})=\sum_{\nu}V_{\nu}({\bf r}-{\bf R}_{\nu})=
\sum_{\nu}\frac{2\pi\hbar^2}{m_{\nu}}a_{\nu}
\delta({\bf r}-{\bf R}_{\nu}).
\eeq
Here ${\bf r}$ and ${\bf R}_{\nu}$ are the position vectors for
neutron with mass $m$ and $\nu$-th nucleus with mass $M_{\nu}$,
$m_{\nu }=mM_{\nu}/(m+M_{\nu})$ is their reduced mass, $a_{\nu}$
is the amplitude of neutron scattering on free nucleus, connected
with the scattering length
$\alpha_{\nu }$ and impact momentum in the center-of-mass system
$k_{\nu }$, in linear on $k_{\nu }\alpha _{\nu }$ approximation
(valid for slow neutrons), by
\beq{1.2}
a_{\nu}=\alpha_{\nu}(1-ik_{\nu}\alpha_{\nu}).
\eeq

For thermal and cold neutrons rescattering of se\-con\-da\-ry
waves is unimportant and one may use Born approximation that gives
for double differential cross section per one target nucleus
\beq{1.3}
\frac{d^2\sigma}{d\Omega d\omega}=
\frac{k'}{2\pi Nk}\sum_{\nu\nu'}
b_{\nu}^*b_{\nu'}\chi(\nu\nu',\vkappa,\omega).
\eeq
Here $\chi (\nu \nu',\vkappa,\omega)$ is the Fourier transform
\beq{1.4}
\chi(\nu\nu',\vkappa,\omega)=
\int\limits_{-\infty}^{+\infty}
\chi(\nu\nu',\vkappa,t)e^{i\omega t}dt
\eeq
of correlation function
\beq{1.5}
\chi(\nu\nu',\vkappa,t)=
\av{i|e^{-i\vkappa\hat{\bf R}_{\nu}(t)}
e^{i\vkappa\hat{\bf R}_{\nu'}(0)}|i},
\eeq
$\vkappa={\bf k}-{\bf k}'$ and $\omega=\epsilon-\epsilon'$ are the
neutron momentum and energy transfers. The quantity
$b_{\nu}=(m/m_{\nu})a_{\nu}$ is called scattering amplitude on
bound nucleus, and $\hat{\bf R}_{\nu}(t)$ is time dependent
Heisenberg operator of nuclear position.

It is easy to show that rescattering of the secondary waves may be
indeed neglected for thermal and cold neutrons. Coherent summation
of the secondary waves may be formed only from the volume
$\sim\lambda^{3}.$ The sum of all amplitudes from this domain is
by the order of magnitude
$\sim n\lambda^{3}(b/\lambda)$, where $n$ is the density of the
nuclei ($\sim 10^{22}$~cm$^{-3}$), and $b\sim 10^{-12}$~cm. So far
as
\beq{1.6}
nb\lambda^2\ll 1,\quad\mbox{i.e.}\quad
\lambda\ll 100~\mbox{nm},
\eeq
rescattering can be neglected. For ultracold neutrons (UCN), when
$\lambda\geq 10$~nm, rescattering of neutron wave in media is very
essential, and for the neutron wave vector $k$ which satisfies
$k^{2}<4\pi nb$, rescattering becomes the dominant process and
results in the total
reflection from the surface of the target (of cause, for positive
$b$). Thus, Born approximation in general, and the cross section
(\ref{1.3}) in particular, can not be used for UCN.

If the Born approximation is unjustified then one
should start from an exact Schr\"{o}dinger equation for the
scattering problem. As the first step one may use a target model
with fixed (unmovable) nuclei and consider integral equation
\beq{1.7}
\Psi_{{\bf k}}({\bf r})=e^{i{\bf k}{\bf r}}-
\frac{m}{2\pi\hbar^2}\int\frac{e^{ik|{\bf r}-{\bf r}'|}}
{|{\bf r}-{\bf r}'|}V({\bf r}')
\Psi_{{\bf k}}({\bf r}')d^3r'.
\eeq
With a formal use of the Fermi pseudo-potential (\ref{1.1})
equation (\ref{1.7}) transforms into
\beq{1.8}
\Psi_{{\bf k}}({\bf r})=e^{i{\bf k}{\bf r}}-
\sum_{\nu}b_{\nu}\frac{e^{ik|{\bf r}-{\bf R}_{\nu}|}}
{|{\bf r}-{\bf R}_{\nu}|}
\Psi_{{\bf k}}({\bf R}_{\nu}).
\eeq

The quantity $\Psi_{{\bf k}}({\bf R}_{\nu })$ seems to have the
meaning of the neutron wave amplitude on the $\nu$-th nucleus.
This wave is composed now from the incident wave and all reflected
waves, so its value is not known in advance and consistent
equations have to be formulated
for these quantities. It is impossible to get the mentioned
equation from (\ref{1.8}) just by substitution there for neutron
position ${\bf r}={\bf R}_{\nu}$, due to infinity in diagonal term
of the right-hand side. So, it was in fact postulated that a
proper equation may be obtained just by throwing away the diagonal
term (self-scattering):
\beq{1.9}
\Psi_{{\bf k}}({\bf R}_{\nu})=e^{i{\bf k}{\bf R}_{\nu}}-
\sum_{\nu'\ne\nu}b_{\nu'}
\frac{e^{ik|{\bf R}_{\nu}-{\bf R}_{\nu'}|}}
{|{\bf R}_{\nu}-{\bf R}_{\nu'}|}
\Psi_{{\bf k}}({\bf R}_{\nu'}).
\eeq

The equations (\ref{1.8}) and (\ref{1.9}) are today the basis for
the whole theory of ultracold neutron interaction with matter
(see, e.g., \cite{Ste77,Gol79,Ign90,Gol91}). From (\ref{1.9}) one
can get an effective repulsive (optical) potential on the
condensed matter surface, so the neutron wave with the energy
below the threshold is exponentially decreasing deep into target.

In general the problem of neutron interaction with matter can be
analyzed in the framework of Multiple Scattering Theory (MST)
\cite{Wat57,Gol64,Sea78,Sea82}. MST deals with interaction of a
projectile with many-body target. In this theory a formal solution
of many-body problem takes the form
\beq{1.10}
\Psi^{(+)}=\Psi^{0}+\hat D^{-1}\sum_{\nu}\hat t_{\nu}\Psi_{\nu},
\eeq
with functions $\Psi_{\nu}$ defined by the set of linear equations
\beq{1.11}
\Psi_{\nu}=\Psi^{0}+\hat D^{-1}\sum_{\nu'\ne\nu}
\hat t_{\nu'}\Psi_{\nu'},
\eeq
where operators $\hat t_{\nu}$ ($t$-matrixes) are linked with
potentials $\hat V_{\nu}$ by
\beq{1.12}
\hat t_{\nu}=\hat V_{\nu}+\hat V_{\nu}\hat D^{-1}\hat t_{\nu}.
\eeq
Here $\Psi^{0}$ is the wave function for noninteracting projectile
and target, and $\hat D^{-1}$ is the "Green function" (see details
in the next section).

There is a wide spread opinion in literature that (\ref{1.8}) and
(\ref{1.9}) are just the equations of MST (\ref{1.10}) and
(\ref{1.11}) for fixed nuclei target. In fact, they have similar
structure but with some modeled $t_{\nu}$. Consistent derivation
of (\ref{1.10}) and (\ref{1.11}) analogies for fixed nuclei target
with realistic potentials was done in \cite{Bar99}. In any way,
target model with fixed nuclei can be applied only to elastic
scattering. While UCN escape from vessels, that is studied for
many years, is due mainly to inelastic scattering. Now
quasielastic scattering of UCN with small energy and momentum
transfers attracts attention of experimenters as perspective tool
for condensed matter studies \cite{Gol96}. Recently observed (see,
e.g., \cite{Bon98,Nes99}) small cooling and heating of UCN in
vessels also belong to quasielastic processes.

A step from frozen to moving nuclei target is very dramatic since
it requires a transition from one body to many-body function
$\Psi({\bf r},{\bf R}_{\nu})$. MST does not present any universal
solution, since general equations of MST are only a reformulation
of the problem in the way where multiple scattering is clearly
exhibited (by using iterated (\ref{1.11}) in (\ref{1.10})).
Practical content of MST is, in fact, a set of approximations
applicable for different situations. They were analyzed, e.g., in
monograph of Goldberger and Watson \cite{Gol64}.

Attempts were made to consider inelastic processes for UCN by
using one of approaches developed in MST (see, e.g.,
\cite{Ste76}). However, approximations used so far for MST cannot
be applied for inelastic scattering of UCN. Indeed, the main
assumptions, that different approximations of MST were based on,
may be formulated as:

(a)~energy of projectile is much larger then characteristic energy
of target particles ("weak coupling approximation");

(b)~mean free path of projectile in target is much larger then its
wave length;

(c)~mean free path of projectile is much larger then length of
effective correlation between target particles.

The first of these assumptions allows to use Born or "impulse"
approximation, where each target particle may be considered free
when colliding with projectile. The second and the third condition
allow to treat multiple scattering as sequential collisions and
represent result as sum on number of collisions executed. Due to
(b) the energy of projectile between successive collisions is well
determined quantity, and due to (c) the target at each collision
may be considered being in the ground state.

It is easy to see that all three assumptions are not valid for
UCN.

(a)~Energy of ultracold neutron is $\sim 10^{-7}$~eV and
corresponds to temperature
\mbox{$\sim 10^{-3}$~K} which is much smaller then the target
temperature even at liquid helium.

(b)~The usual definition of mean free path fails for UCN (elastic
cross section for mirror like potential is equal to surface area
S). Thus, one may use for this quantity intrusion length into
target, which is of the order of wave length $\sim 10$~nm. It
means that neutron energy between collisions is, in fact,
uncertain.

(c)~The main effect of UCN multiple scattering is appearance of a
potential barrier, that is just the product of particle-particle
correlation at distances compared with neutron wave length.

The goal of this work is to find solution of MST equations valid
for the whole domain of slow neutrons (from thermal to ultracold),
starting from realistic neutron -- nucleus interaction. Equations
(\ref{1.3}) and (\ref{1.8}), (\ref{1.9}) will follow from this
theory as limiting cases. By solution we mean reduction of general
equations (\ref{1.10})-(\ref{1.12}) to those which allow
reasonably simple numerical solution for elastic and inelastic
scattering for all practically interesting cases. One numerical
solution for inelastic UCN scattering is presented as an example
in the last part of the paper.

\section{Formulation of the Problem. Plan of Solution}

A proper theory for UCN scattering should be based on the
following postulates: (i) No Born approximation; (ii) No use of
Fermi potential; (iii) Target matter is a dynamical system. So we
should start from N+1 body Schr\"{o}dinger equation
\beq{2.1}
\left(\frac{\hat{\bf p}^2}{2m}+\hat H_t+\hat V\right)
|\Psi_{{\bf k},i}\rangle=E_{{\bf k},i}|
\Psi_{{\bf k},i}\rangle,\quad
\hat V=\sum_{\nu}\hat V_{\nu}.
\eeq
Here $\hat V_{\nu }$ describes interaction of neutron with
$\nu$-th nucleus,
$\hat{\bf p}=-i\partial/\partial {\bf r}$ is the operator of
neutron momentum,
$E_{{\bf k},i}=\epsilon_{{\bf k}}+\varepsilon_i$ is the total
energy as the sum of neutron energy $\epsilon _{{\bf k}}=k^2/2m$
in the state $|{\bf k}\rangle$ and the target initial energy
$\varepsilon _i$ in the state $|i\rangle$ that is the eigenstate
of the target Hamiltonian $\hat H_t$. Here and onward we keep
$\hbar =1$ till final physical results.

Equation (\ref{2.1}) can be written in integral form
\beq{2.2}
|\Psi_{{\bf k},i}\rangle=|\Psi^0_{{\bf k},i}\rangle+
\hat D^{-1}\sum_{\nu}\hat V_{\nu}|\Psi_{{\bf k},i}\rangle,
\eeq
where $|\Psi^0_{{\bf k},i}\rangle=|{\bf k}\rangle|i\rangle$ and
$\hat D^{-1}$ is ''Green function'' with
\beq{2.3}
\hat D=\frac{k^2}{2m}+\varepsilon_i-
\frac{\hat{\bf p}^2}{2m}-\hat H_t+i\eta,
\eeq
where positive quantity $\eta\to 0$ provides outgoing neutron wave
asymptotic.

Note, that the problem can be easily reduced to MST equations
(\ref{1.10})-(\ref{1.12}). First, equation (\ref{2.2}) can be
written in the form
\beq{2.3d}
|\Psi_{{\bf k},i}\rangle-
\hat D^{-1}\hat V_{\nu}|\Psi_{{\bf k},i}\rangle=
|\Psi^0_{{\bf k},i}\rangle+
\hat D^{-1}\sum_{\nu'\ne\nu}\hat V_{\nu'}|\Psi_{{\bf k},i}\rangle.
\eeq
Then, let define a state vector $|\Psi_{\nu}\rangle$ and operator
$\hat t_{\nu}$ by the relations
\beq{2.3b}
|\Psi_{\nu}\rangle=|\Psi_{{\bf k},i}\rangle-
\hat D^{-1}\hat V_{\nu}|\Psi_{{\bf k},i}\rangle,
\eeq
\beq{2.3c}
\hat t_{\nu}|\Psi_{\nu}\rangle=
\hat V_{\nu}|\Psi_{{\bf k},i}\rangle.
\eeq
Now one can see that equations (\ref{2.2}) and (\ref{2.3d}) with
the help of (\ref{2.3b}) and (\ref{2.3c}) turns into (\ref{1.10})
and (\ref{1.11}), respectively. Finally, we have to show that
$\hat t_{\nu}$ obeys (\ref{1.12}). For this purpose we use
identity
\beq{2.3e}
\hat V_{\nu}|\Psi_{{\bf k},i}\rangle=
\hat V_{\nu}(|\Psi_{{\bf k},i}\rangle-
\hat D^{-1}\hat V_{\nu}|\Psi_{{\bf k},i}\rangle)+
\hat V_{\nu}\hat D^{-1}\hat V_{\nu}|\Psi_{{\bf k},i}\rangle,
\eeq
that, with the help of (\ref{2.3b}), (\ref{2.3c}), transforms into
\beq{2.3f}
\hat t_{\nu}|\Psi_{\nu}\rangle=
\hat V_{\nu}|\Psi_{\nu}\rangle+
\hat V_{\nu}\hat D^{-1}\hat t_{\nu}|\Psi_{\nu}\rangle.
\eeq
Thus we have demonstrated that MST equations are
nothing more than re\-for\-mu\-la\-tion of the ge\-ne\-ral
scat\-te\-ring prob\-lem (\ref{2.1}) or (\ref{2.2}).

It is, of course, impossible to solve the many-body equation
(\ref{2.1}) or MST equations (\ref{1.10})-(\ref{1.12}) and to
found the state vectors $|\Psi_{{\bf k},i}\rangle$ or
$|\Psi_{\nu}\rangle$ without any approximations. In our problem
there are two main small parameters: short-range of
neutron -- target nuclei interaction (as compared with interatomic
distance and wave length) and small neutron energy (as compared
with depth of interaction potential).

The first condition allows to consider only s-wave part of the
wave function of neutron -- nucleus center-of-mass motion, when
their interaction is evaluated. And the second condition allows in
this evaluation to neglect energy of relative neutron -- nucleus
motion inside the interaction potential area. So, the s-wave
function and its derivative, taken at the potential boundary, are
independent on neutron energy and are just numerical parameters.

No specific model for neutron -- nucleus interaction potential
will be needed. Its specific features described above (short range
and large depth) allows to use scattering length approximation.

The small parameters allow to simplify our problem. Potential
$V_{\nu}$ is essential only in small vicinity of ${\bf R}_{\nu}$.
It differs from zero only when the absolute value of deviation
${\bf x}={\bf r}-{\bf R}_{\nu}$ does not exceed the potential
radius $r_{0\nu}$. So, using completeness of the neutron states
$\sum_{{\bf r}}|{\bf r}\rangle\langle{\bf r}|=1$ one has
\beq{2.3g}
\begin{array}{l}
\hat V_{\nu}|\Psi_{{\bf k},i}\rangle=
\sum_{{\bf x}}|{\bf R}_{\nu}+{\bf x}\rangle
\av{{\bf R}_{\nu}+{\bf x}|\hat V_{\nu}|\Psi_{{\bf k},i}}={}
\\[\medskipamount]
\phantom{\hat V_{\nu}|\Psi_{{\bf k},i}\rangle}=
\sum_{{\bf x}}|{\bf R}_{\nu}+{\bf x}\rangle V_{\nu}({\bf x})
\av{{\bf R}_{\nu}+{\bf x}|\Psi_{{\bf k},i}}.
\end{array}
\eeq
Here and onward sums of continue variables mean integrals with the
following supposition for position- and momentum--energy variables
\beq{2.4}
\sum_{{\bf R}}\longrightarrow \int d\,{\bf R},\quad
\sum_{{\bf q}}\longrightarrow
\int\frac{d\,{\bf q}}{(2\pi) ^3},\quad
\sum_{\omega}\longrightarrow \int\frac{d\,\omega}{2\pi}.
\eeq

Using (\ref{2.3g}) one can transform (\ref{2.2}) to
\beq{2.5}
|\Psi_{{\bf k},i}\rangle=
|\Psi^0_{{\bf k},i}\rangle+
\hat D^{-1}\sum_{\nu}\sum_{{\bf x}}
|{\bf R}_{\nu}+{\bf x} \rangle V_{\nu}({\bf x})
\av{{\bf R}_{\nu}+{\bf x}|\Psi_{{\bf k},i}}.
\eeq
The scalar product of (\ref{2.3d}) with
$\langle {\bf R}_{\nu}+{\bf x}|$ after some rearrangement can be
presented as
\beq{2.5b}
\begin{array}{l}
\av{{\bf R}_{\nu}+{\bf x}|\Psi_{{\bf k},i}}-{}
\\[\medskipamount]
\phantom{}-
{\ds\sum_{{\bf x}'}}\av{{\bf R}_{\nu}+{\bf x}|\hat D^{-1}|
{\bf R}_{\nu}+{\bf x}'}V_{\nu}({\bf x}')
\av{{\bf R}_{\nu}+{\bf x}'|\Psi_{{\bf k},i}}={}
\\[\medskipamount]
\phantom{}=
\av{{\bf R}_{\nu}+{\bf x}|\Psi^0_{{\bf k},i}}+{}
\\[\medskipamount]
\phantom{}+
{\ds\sum_{\nu'\ne\nu}\sum_{{\bf x}'}}
\av{{\bf R}_{\nu}+{\bf x}|\hat D^{-1}|
{\bf R}_{\nu'}+{\bf x}'}V_{\nu'}({\bf x}')
\av{{\bf R}_{\nu'}+{\bf x}'|\Psi_{{\bf k},i}}.
\end{array}
\eeq
Equations (\ref{2.5}) and (\ref{2.5b}), as can be recognized by
the structures of their right-hand sides, correspond to MST
equations (\ref{1.10}) and (\ref{1.11}). To make them fully
determined it remains to find the only key element, namely
$\av{{\bf R}_{\nu}+{\bf x}|\Psi_{{\bf k},i}}$, i.e. the exact
many-body wave function, but only in the area of short range
potential for each nucleus.

So, the many-body problem is reduced to only two-body problem (or,
more precisely, three-body, since the nucleus is not free), with
the rest of the nuclei as spectators. The solution of the last
problem is determined by two parameters: by the values of s-wave
amplitude
$\chi_{\nu }(r_{0\nu})/r_{0\nu}$ and its derivative at the
boundary of the potential. Since the logarithmic derivative is
connected to the scattering length $\alpha_{\nu}$ (the known
physical quantity), the amplitude $\chi_{\nu}(r_{0\nu})$ remains
the only free parameter.

Therefore one may hope to express the quantity
$\av{{\bf R}_{\nu}+{\bf x}|\Psi_{{\bf k},i}}$ by
$\chi_{\nu}(r_{0\nu})$ and then to use (\ref{2.5b}) as a set of
linear equations for parameters $\chi_{\nu}(r_{0\nu})$. We will
call
$\av{{\bf R}_{\nu}+{\bf x}|\Psi_{{\bf k},i}}$ for $x=r_{0\nu}$ as
"neutron function at the surface of  $\nu$-th nucleus".

{\it Remark}. The small quantity $r_{0\nu}$ will be neglected
whenever possible, except in cases where $r_{0\nu}$ stays near the
scattering length (which may be of the same order of magnitude)
and in terms with singularity $1/r_{0\nu}$ till their
compensation. Such singularity occurs in both terms in the
left-hand side of (\ref{2.5b}) and requirement of their
compensation will give us an additional control of calculations.

\section{Neutron function at the surface of $\nu$-th nucleus}

Let us transform the basic equation (\ref{2.1}) to new variables
\beq{3.1}
{\bf r},\,{\bf R}\quad\longrightarrow\quad
{\bf x}={\bf r}-{\bf R}_{\nu},\,{\bf R},
\eeq
where ${\bf R}=\{{\bf R}_{\nu}\}$. The Hamiltonian in the new
variables takes the form
\beq{3.2}
\hat H=\hat H_n+\hat H_t-
\frac{\hat{\bf p}_x\hat{\bf P}_{\nu}}{M_{\nu}},\quad
\hat H_n=\frac{\hat{\bf p}_x^2}{2m_{\nu}}+V_{\nu}(x),
\eeq
where $\hat{\bf p}_x=-i\partial/\partial{\bf x}$, and
$\hat{\bf P}_{\nu}=-i\partial/\partial{\bf R}_{\nu}$ is the
momentum operator of $\nu$-th nucleus. We assume that the quantity
${\bf x}$ is small, so that neutron interaction with other nuclei
(with $\nu'\neq\nu $) is absent.

We look for the solution of the Schr\"{o}dinger equation with the
Hamiltonian (\ref{3.2}) with the energy $E_{{\bf k},i}={}$
${\bf k}^2/2m+\varepsilon_i$. In Born approximation in the limit
$x\to 0$ solution has the form
\beq{3.3}
\Psi_{{\bf k,}i}({\bf x},{\bf R})=\psi({\bf x})
e^{i{\bf k}{\bf R}_{\nu }}\Phi_i({\bf R}),
\eeq
where $\psi({\bf x})$ is the center-of-mass wave function, and
$\Phi_i({\bf R})$ is the initial target state vector. In general
case the neutron near the $\nu$-th nucleus may have its energy
different from the initial one due to previous collisions. So, it
is natural to look for the solution in the form
\beq{3.4}
\Psi_{{\bf k,}i}({\bf x},{\bf R})=\varphi({\bf x})
e^{i{\bf g}{\bf R}_{\nu }}\Phi_j({\bf R}),
\eeq
where $\Phi_j({\bf R})$ is the eigenfunction of the target
Hamiltonian $H_t$ with the energy $\varepsilon_j$ and ${\bf g}$ is
a vector parameter which represents neutron momentum.

If we substitute (\ref{3.4}) into Schr\"{o}dinger equation with
the Hamiltonian (\ref{3.2}) and take into account that operator
$\hat{\bf P}_{\nu}$ acts on $\Phi_j({\bf R})$ as well as on the
exponent, then we obtain the following equation for
$\varphi({\bf x})$
\beq{3.5}
\left(\hat H_n+
\frac{{\bf g}^2}{2M_{\nu}}+
\frac{{\bf g}{\bf G}_{j\nu}}{M_{\nu}}\right)\varphi-
\frac{({\bf g}+{\bf G}_{j\nu})\hat{\bf p}_x}{M_{\nu}}\varphi=
(E_{{\bf k},i}-\varepsilon_j)\varphi,
\eeq
where
\beq{3.6}
{\bf G}_{j\nu}=
\left(\hat{\bf P}_{\nu}\Phi_j\right)/\Phi_j.
\eeq

Equation (\ref{3.5}) after some formal transformation can be
displayed as
\beq{3.7}
\left(\frac{1}{2m_{\nu }}\left[\hat{\bf p}_x-
\frac{m_{\nu}}{M_{\nu }}({\bf g}+{\bf G}_{j\nu})\right]^2+
V_{\nu}(x)\right)\varphi =E({\bf R})\varphi,
\eeq
where
\beq{3.8}
E({\bf R})=E_{{\bf k},i}-\varepsilon_j-
\frac{g^2}{2m}+\frac{1}{2m_{\nu }}
\left[{\bf g}-\frac{m_{\nu }}{M_{\nu }}({\bf g}+
{\bf G}_{j\nu})\right]^2.
\eeq

Formally, (\ref{3.7}) is an equation only for the function
$\varphi$, and its exact solution, as can be easily proved, may be
presented in the form
\beq{3.9}
\varphi({\bf x})=\exp\left(i\Frac{m_{\nu}}{M_{\nu}}
({\bf g}+{\bf G}_{j\nu}){\bf x}\right)
\Psi({\bf q},{\bf x}),
\eeq
where $\Psi({\bf q},{\bf x})$ is the scattering wave-function in
the cen\-ter-of-mass system determined by the equation
\beq{3.10}
\hat H_n\Psi({\bf q},{\bf x})=
\frac{q^2}{2m_{\nu}}\Psi({\bf q},{\bf x}),
\eeq
and the scattering energy is defined from
\beq{3.11}
\frac{q^2}{2m_{\nu}}=
\varepsilon_i-\varepsilon_j+\frac{k^2-g^2}{2m}+
\frac{1}{2m_{\nu}}
\left[{\bf g}-\frac{m_{\nu }}{M_{\nu }}({\bf g}+
{\bf G}_{j\nu})\right]^2.
\eeq

So, initial problem of neutron scattering on the bound nucleus
seems to be reduced to the problem of scattering on free nucleus.
In fact, it cannot be done precisely. Indeed, though (\ref{3.9})
is formally an exact solution of the equation (\ref{3.5}), its
parameter ${\bf q}$ -- impact momentum -- through vector ${\bf
G}_{j\nu}$ (\ref{3.6}) depends on all coordinates of the target
${\bf R}$. But such a dependance was not assumed in (\ref{3.4}).

However, we are looking for the solution valid for small $x$. In
the limit of small $x$ only the $s$-wave part $\chi_{\nu}(x)/x$ of
the scattering function $\Psi({\bf q},{\bf x})$ is of importance
and dependance of $\chi_{\nu}$ on $q$ can be neglected. Therefore,
expression (\ref{3.9}) for small $x$ is independent on ${\bf R}$
and gives the real solution in the form
\beq{3.12}
\Psi_{{\bf k},i}({\bf x},{\bf R})_{|x\to 0}\simeq
\frac{\chi_{\nu}(x)}{x}
e^{i{\bf g}{\bf R}_{\nu}}\Phi_j({\bf R}).
\eeq

So far the target state $j$ and intermediate neutron momentum
${\bf g}$ were not specified. It is evident, that any linear
combination of functions (\ref{3.12}) is allowed (provided the
right-hand side of (\ref{3.11}) is non negative), so we finally
obtain
\beq{3.13}
\Psi_{{\bf k},i}({\bf r}=
{\bf R}_{\nu}+{\bf x},{\bf R})_{|x\to 0}\simeq
\frac{\chi_{\nu}(x)}{x}e^{i{\bf k}{\bf R}_{\nu}}\av{{\bf R}|\nu},
\eeq
where
\beq{3.14}
\av{{\bf R}|\nu}\equiv\Phi_{\nu}({\bf R})=
\sum_{{\bf g},j}C_{\nu}({\bf g},j)
e^{i({\bf g}-{\bf k}){\bf R}_{\nu}}\Phi_j({\bf R}).
\eeq
Note, that $C_{\nu}({\bf g},j)=\delta_{{\bf g}{\bf k}}\delta_{ji}$
and $\Phi_{\nu}\to\Phi_i$ in Born approximation.

Expression (\ref{3.13}) defines a local structure of the basic
function $\Psi _{{\bf k},i}$ near the point
${\bf r}={\bf R}_{\nu}$. It naturally contains a "background
target function" $\Phi _{\nu }({\bf R}) =\av{{\bf R}|\nu}$,
separately defined for each nucleus. This function differs from
the initial state of the target
$\Phi_i({\bf R})\equiv\av{{\bf R}|i}$ due to perturbation by the
neutron wave. The functions $\av{{\bf R}|\nu}$ should be
consistently determined in parallel to the amplitudes of the
neutron wave $\chi_{\nu }(r_{0\nu})$.

Let us now calculate the integral over ${\bf x}$ in (\ref{2.3g}).
First, from (\ref{3.13}) one finds
\beq{3.15}
\av{{\bf R}_{\nu}+{\bf x}|\Psi_{{\bf k},i}}\simeq
\frac{\chi_{\nu}(x)}{x}e^{i{\bf k}{\bf R}_{\nu}}|\nu\rangle.
\eeq
Second, we note that the state vector
$|{\bf R}_{\nu}+{\bf x}\rangle$ due to smooth dependance on
${\bf x}$ may be factored out from the integral and taken at
$x=0$. Third, from the equation (\ref{3.10}) for $\chi_{\nu}$
inside the potential area it follows
\beq{3.16}
V_{\nu}(x)\chi_{\nu}(x)\simeq
\frac{1}{2m_{\nu}}\,
\frac{d^2\chi_{\nu}(x)}{dx^2}.
\eeq
Thus, the integration can be performed
\beq{3.17}
\int V_{\nu }(x)\frac{\chi _{\nu }(x)}{x}d{\bf x}=
-\frac{2\pi}{m_{\nu}}\,
\frac{\alpha_{\nu}\chi_{\nu}(r_{0\nu})}
{\alpha_{\nu}-r_{0\nu}},
\eeq
where we used the relation between scattering length $\alpha$ and
the logarithmic derivative at the boundary of the potential
$[\chi'/\chi] _{r=r_0}=-1/(\alpha-r_0)$. Combining (\ref{3.15})
and (\ref{3.17}) we get
\beq{3.18}
\sum_{{\bf x}}|{\bf R}_{\nu}+{\bf x}\rangle
V_{\nu }({\bf x})\av{{\bf R}_{\nu}+{\bf x}|\Psi_{{\bf k},i}}=
|{\bf R}_{\nu}\rangle
\frac{2\pi }{m}e^{i{\bf k}{\bf R}_{\nu }}|\nu\rangle\phi_{\nu},
\eeq
where we have introduced the amplitude
\beq{3.19}
\phi_{\nu}=-\frac{m}{m_{\nu}}\,
\frac{\alpha_{\nu}\chi_{\nu}(r_{0\nu})}{\alpha_{\nu}-r_{0\nu}}.
\eeq

\section{Equations for neutron amplitudes}

In the previous section integrals over ${\bf x}$ in (\ref{2.5b})
are expressed in terms of the amplitudes $\phi_{\nu}$ and state
vectors $|\nu\rangle$. Thus, with the use of (\ref{3.15}) the
equation (\ref{2.5b}) takes the form
\beq{5.1}
\begin{array}{l}
\Frac{\chi_{\nu}(x)}{x}e^{i{\bf k}{\bf R}_{\nu}}|\nu\rangle-
\av{{\bf R}_{\nu}+{\bf x}|\hat D^{-1}|{\bf R}_{\nu}}
\Frac{2\pi }{m}e^{i{\bf k}{\bf R}_{\nu}}|\nu\rangle\phi_{\nu}={}
\\[\medskipamount]
\phantom{}=
e^{i{\bf k}{\bf R}_{\nu}}|i\rangle+
{\ds\sum_{\nu'\ne\nu}}\av{{\bf R}_{\nu}|\hat D^{-1}|
{\bf R}_{\nu'}}
\Frac{2\pi }{m}e^{i{\bf k}{\bf R}_{\nu'}}|\nu'\rangle\phi_{\nu'}.
\end{array}
\eeq
The small quantity $x$ is left only in the terms with singularity
$1/x$ and neglected elsewhere.

Note, that $x$ in (\ref{5.1}) is a free parameter, and for
$x=r_{0\nu}$, when $\chi_{\nu}(r_{0\nu})$ and $\phi_{\nu}$ are
linked by (\ref{3.19}), relation (\ref{5.1}) may be regarded as a
set of linear equations for $\phi_{\nu}$, but with operators in
target states as coefficients could be hardly used for practical
calculations.  Our next steps are directed to simplification of
(\ref{5.1}).

It is useful to transform the terms with $\hat D^{-1}$ as follows
\beq{5.2}
\av{{\bf R}|\hat D^{-1}|{\bf R}'}=
\sum_{{\bf q}}e^{i{\bf q}{\bf R}}\hat D_q^{-1}
e^{-i{\bf q}{\bf R}'},
\eeq
where
\beq{5.3}
\hat D_q=\av{{\bf q}|\hat D|{\bf q}}=
\frac{k^2-q^2}{2m}+\varepsilon_i-\hat H_t+i\eta.
\eeq
Then from (\ref{5.1}) it is evident that to separate the
amplitudes one should multiply both sides by
$e^{-i{\bf k}{\bf R}_{\nu}}$ and take their scalar product with
the eigenvector $\langle j|$ of
$\hat H_t$. Thus, we have
\beq{5.4}
\begin{array}{l}
\left[\Frac{m_{\nu }}{m}
\left(\Frac{1}{\alpha_{\nu }}-\Frac{1}{r_{0\nu}}\right)
\av{j|\nu}-
\zeta_j(\nu\nu,{\bf x})_{|x\to r_{0\nu}}\right]\phi_{\nu}={}
\\[\medskipamount]
\phantom{\Frac{m_{\nu }}{m}
\left(\Frac{1}{\alpha_{\nu }}-
\Frac{1}{r_{0\nu}}\right)}=\delta_{ij}+
{\ds\sum_{\nu'\ne\nu}}\zeta_j(\nu\nu',0)\,\phi_{\nu'},
\end{array}
\eeq
where
\beq{5.5}
\zeta_j(\nu\nu',{\bf x})=\frac{2\pi}{m}
\sum_{{\bf q}}e^{i{\bf q}{\bf x}}
\av{j|e^{-i({\bf k}-{\bf q})\hat{\bf R}_{\nu }}\hat D_q^{-1}
e^{i({\bf k}-{\bf q})\hat{\bf R}_{\nu'}}|\nu'}.
\eeq

Expression (\ref{5.5}) can be transformed into the other form with
the use of the following presentation of the operator
$\hat D_q^{-1}$
\beq{5.6}
\hat D_q^{-1}=\frac{1}{i}\int\limits_0^{\infty }
e^{ i(\epsilon_{{\bf k}}-\epsilon_{{\bf q}}+
\varepsilon _i-\hat H_t+i\eta)t}\,dt.
\eeq
Then we get the alternative expression for (\ref{5.5})
\beq{5.7}
\begin{array}{l}
\zeta_j(\nu\nu',{\bf x})=\Frac{2\pi }{im}\times{}
\\[\bigskipamount]
\phantom{}\times{\ds\sum_{{\bf q}}}
e^{i{\bf q}{\bf x}}{\ds\int\limits_0^{\infty}}
\chi_j(\nu\nu',{\bf k}-{\bf q},t)\,
e^{i(\epsilon_{{\bf k}}-\epsilon_{{\bf q}}+
\varepsilon_i-\varepsilon_j+i\eta)t}\,dt.
\end{array}
\eeq
Here the correlation function is introduced
\beq{5.7b}
\chi_j(\nu\nu',\vkappa,t)=\av{j|e^{-i\vkappa\hat{\bf R}_{\nu}(t)}
e^{i\vkappa\hat{\bf R}_{\nu'}(0)}|\nu'},
\eeq
which is a generalization of (\ref{1.5}).

To make (\ref{5.4}) fully determined, we only need to find an
explicit expression for diagonal term $\zeta_j(\nu\nu,{\bf x})$ in
the limit $x\to r_{0\nu}$. It evidently has a singularity
$1/r_{0\nu}$ which should compensate analogous term on the
left-hand side of (\ref{5.4}). That can be easily seen from simple
scaling analysis. Let introduce new variables and parameters
\beq{5.8}
{\bf q}'={\bf q}x,\quad
{\bf k}'={\bf k}x,\quad
\varepsilon'_i-\varepsilon'_j=
(\varepsilon_i-\varepsilon_j)x^2,\quad
t'=t/x^2.
\eeq
Then small parameter $x$ remains in (\ref{5.7}) only as the common
factor $1/x$ and inside $\chi_j$ in the argument of
${\bf R}_{\nu}(t'x^2)$.

When $\nu=\nu'$ the operator in matrix element (\ref{5.7b}) is
close to unity with very small deviation
$\sim u\kappa$, where $u$ is a nuclear shift from equilibrium.
Thus, it seems to be a good approximation to integrate over $t$ in
(\ref{5.7}) only the exponential factor outside the matrix
element. Then the integration over ${\bf q}$ gives
\beq{5.9}
\zeta_j(\nu\nu,{\bf x})\simeq
-\frac{e^{ik_jx}}{x}\av{j|\nu},
\eeq
where
\beq{5.9b}
k_j^2=k^2+2m(\varepsilon_i-\varepsilon_j).
\eeq
Note, that $k_j$ is the absolute value of momentum of
inelastically scattered neutron when target remains in the
eigenstate $|j\rangle$.

However, this result is evidently wrong. Since it does not cancel
the singularity $1/r_{0\nu}$ in the left-hand side of (\ref{5.4})
due to the additional factor $m_{\nu}/m$. So, the matrix element
$\chi_j$ should be more accurately estimated.

The limit $x\to 0\ $means that only small time interval is of
importance in
${\bf R}_{\nu }(t)={\bf R}_{\nu}(t'x^2)$, and we can use the
expansion
\beq{5.10}
\hat{\bf R}_{\nu }(t)\simeq\hat{\bf R}_{\nu }(0)+
\frac{\hat{\bf P}_{\nu}t}{M_{\nu }}.
\eeq
Each additional power in $t$ will result in an additional factor
$x^2$. After substitution of (\ref{5.10}) into (\ref{5.7b}), the
matrix element is calculated without any father approximations.
Using the identity
$\exp(\hat A)\exp(\hat B)=\exp(\hat A+\hat B+[\hat A,\hat B]/2)$,
which holds when $\hat A$ and $\hat B$ commute with their
commutator $[\hat A,\hat B]$, we obtain
\beq{5.11}
\begin{array}{l}
\chi_j(\nu\nu,{\bf k}-{\bf q},t)_{|t\to 0}\longrightarrow{}
\\[\medskipamount]
\phantom{}\longrightarrow
\av{j|\exp\left(-i\left[
\Frac{({\bf k}-{\bf q})\hat{\bf P}_{\nu }}{M_{\nu }}+
\Frac{({\bf k}-{\bf q})^2}{2M_{\nu}}\right]t\right)|\nu}.
\end{array}
\eeq

Then integration over time and elementary transformation give
\beq{5.13}
\begin{array}{l}
\zeta_j(\nu\nu,{\bf x}) _{|x\to 0}={}
\\[\medskipamount]
\phantom{}=
4\pi\Frac{m_{\nu }}{m}{\ds\sum_{{\bf q}}}
\av{j|\Frac{e^{i{\bf q}{\bf x}}}
{k_{\nu }^2-q_{\nu }^2+
2m_{\nu }(\varepsilon_i-\varepsilon_j) +i\eta}|\nu},
\end{array}
\eeq
where
\beq{5.14}
{\bf k}_{\nu }={\bf k}-\frac{m_{\nu }}{M_{\nu }}
(\hat{\bf {P}}_{\nu }+{\bf k}),\quad
{\bf q}_{\nu }={\bf q}-\frac{m_{\nu }}{M_{\nu }}
(\hat {\bf P}_{\nu }+{\bf k}).
\eeq
The physical meaning of (\ref{5.14}) is evident:
${\bf k}_{\nu}$ and ${\bf q}_{\nu }$ are the momenta in the
center-of-mass system. This transition to the center-of-mass
system provides additional factor $m_{\nu }/m$.

Finally, integrating over ${\bf q}$ and taking into account only
s-wave part in ${\bf x}$ we get
\beq{5.15}
\zeta_j(\nu\nu,{\bf x})_{|x\to 0}=
-\frac{m_{\nu }}{m}\left(\frac{1}{x}\av{j|\nu}+
i\av{j|\hat K_{j\nu}|\nu}\right),
\eeq
where the operator $\hat K_{j\nu }$, defined from
\beq{5.16}
\hat K_{j\nu }^2=\left[{\bf k}-\frac{m_{\nu }}{M_{\nu }}
(\hat{\bf P}_{\nu }+{\bf k})\right]^2+
2m_{\nu }(\varepsilon_i-\varepsilon_j),
\eeq
has the meaning of absolute value of the impact momentum in the
center-of-mass system for the neutron and $\nu$-th target nucleus
when the target is in the state $|j\rangle$. For UCN this quantity
is defined mostly by the average absolute value of nuclear
momentum.

The limiting value (\ref{5.15}) allows to get explicit form of the
equation (\ref{5.4}) for the amplitudes $\phi_{\nu }$. With the
help of (\ref{5.15}) we verify that for $x=r_{0\nu}$ singular
terms $1/r_{0\nu}$ are canceled out and the next terms result in
\beq{5.17}
\frac{1}{\beta _{\nu }}\av{j|\nu}\phi_{\nu }+
i\frac{m_{\nu }}{m}
\av{j|\hat K_{j\nu }|\nu}\phi_{\nu}=
\delta_{ij}+
\sum_{\nu'\ne\nu}
\zeta_j(\nu\nu')\phi_{\nu'},
\eeq
where the renormalized scattering length (''on bound nucleus'') is
introduced
\beq{5.18}
\beta_{\nu}=\frac{m}{m_{\nu }}\alpha_{\nu}.
\eeq
The matrix $\zeta_j(\nu\nu')\equiv\zeta_j(\nu\nu',0)$ with the use
of (\ref{5.7}) can be expressed in terms of Fourier transforms of
correlation functions (\ref{5.7b})
\beq{5.19}
\zeta_j(\nu\nu') =4\pi\sum_{{\bf q},\omega}
\frac{\chi _j(\nu\nu',{\bf k}-{\bf q},\omega)}
{k_j^2-q^2-2m\omega +i\eta}.
\eeq

Now we note that any state vector $|\nu\rangle$ can be presented
as series in target eigenfunctions
\beq{5.19b}
|\nu\rangle=\sum_j|j\rangle\av{j|\nu}.
\eeq
Therefore the relations (\ref{5.17}) are really the linear
equations for the amplitudes
\beq{5.19c}
\phi^j_{\nu}=\av{j|\nu}\phi_{\nu}.
\eeq
Indeed, we get
\beq{5.19d}
\frac{1}{\beta _{\nu }}\phi^j_{\nu }+
i\frac{m_{\nu }}{m}\sum_{j'}
\av{j|\hat K_{j\nu }|j'}\phi^{j'}_{\nu}=
\delta_{ij}+
\sum_{j',\nu'\ne\nu}
\zeta_{jj'}(\nu\nu')\phi^{j'}_{\nu'}.
\eeq
The coefficients of these equations
\beq{5.19e}
\begin{array}{l}
\zeta_{jj'}(\nu\nu')=\Frac{2\pi}{m}
{\ds\sum_{{\bf q}}}
\av{j|e^{-i({\bf k}-{\bf q})\hat{\bf R}_{\nu }}\hat D_q^{-1}
e^{i({\bf k}-{\bf q})\hat{\bf R}_{\nu'}}|j'}={}
\\[\bigskipamount]
\phantom{\zeta_{jj'}(\nu\nu')}=
4\pi{\ds\sum_{{\bf q},\omega}}
\Frac{\chi _{jj'}(\nu\nu',{\bf k}-{\bf q},\omega)}
{k_j^2-q^2-2m\omega +i\eta}
\end{array}
\eeq
with
\beq{5.19f}
\begin{array}{l}
\chi_{jj'}(\nu\nu',\vkappa,\omega)=
\int\limits_{-\infty}^{+\infty}
\chi_{jj'}(\nu\nu',\vkappa,t)e^{i\omega t}dt,
\\[\medskipamount]
\chi_{jj'}(\nu\nu',\vkappa,t)=
\av{j|e^{-i\vkappa\hat{\bf R}_{\nu}(t)}
e^{i\vkappa\hat{\bf R}_{\nu'}(0)}|j'},
\end{array}
\eeq
are completely determined by the properties of target matter,
i.e., by the matrix elements of the operator of $\nu$-th and
$\nu'$-th nuclei position correlation between the target
eigenfunctions.

The equations (\ref{5.19d}) are obtained from (\ref{2.5b}) or
(\ref{1.11}). However, they deal not with N+1 body state vectors
$|\Psi_{{\bf k},i}\rangle$ or $|\Psi_{\nu}\rangle$ but with
numerical parameters $\phi^j_{\nu}$. It is easy to see that due to
(\ref{3.18}) and (\ref{2.5}) the total wave function
$|\Psi_{{\bf k},i}\rangle$ can be expressed in terms of the same
parameters, namely,
\beq{5.19g}
|\Psi_{{\bf k},i}\rangle=|\Psi^0_{{\bf k},i}\rangle+
\hat D^{-1}\sum_{j,\nu}|{\bf R}_{\nu}\rangle
\frac{2\pi }{m}e^{i{\bf k}{\bf R}_{\nu }}|j\rangle\phi^j_{\nu}.
\eeq
It means that scattering probability and scattering cross section
are also determined by $\phi^j_{\nu}$. Connection between them is
analysed in the next section.

\section{Scattering problem and neutron amplitudes}

Scattering process with a fixed final state $\langle j|$ of the
target is usually called a transition into $j$-th reaction
channel. The wave function in the $j$-th channel can naturally be
defined by
\beq{4.6}
\Psi _{ij}({\bf r}) =
\av{{\bf r},j|\Psi _{{\bf k},i}}
\eeq
From (\ref{5.19g}) it follows
\beq{4.7}
\Psi_{ij}({\bf r})=\delta_{ij}e^{i{\bf k}{\bf r}}+
\frac{2\pi }{m}\av{j|\sum_{j',\nu}\phi^{j'}_{\nu}
\av{{\bf r}|\hat D^{-1}|{\bf R}_{\nu}}
e^{i{\bf k}{\bf R}_{\nu}}|j'}.
\eeq
Then using (\ref{5.2}), (\ref{5.3}) one has
\beq{4.7b}
\Psi_{ij}({\bf r})=\delta_{ij}e^{i{\bf k}{\bf r}}+
4\pi\sum_{j',\nu}\phi^{j'}_{\nu}
\av{j|\sum_{{\bf q}}
\frac{e^{i{\bf q}({\bf r}-\hat{\bf R}_{\nu})}}
{k_j^2-q^2+i\eta}e^{i{\bf k}\hat{\bf R}_{\nu}}|j'},
\eeq
where $k_j$ was introduced by (\ref{5.9b}). After integration over
${\bf q}$ we finally obtain
\beq{4.8}
\Psi_{ij}({\bf r})=\delta_{ij}e^{i{\bf k}{\bf r}}-
\sum_{j',\nu}\phi^{j'}_{\nu}
\av{j|\frac{e^{ik_j|{\bf r}-\hat{\bf R}_{\nu }|}}
{|{\bf r}-\hat{\bf R}_{\nu }|}
e^{i{\bf k}\hat{\bf R}_{\nu }}|j'}.
\eeq

In asymptotic, when $r\to\infty$, we get from (\ref{4.8}) for the
scattered wave the usual structure
\beq{4.10}
f_{ij}({\bf k},{\bf k}_j)\frac{e^{ik_jr}}{r},
\eeq
where the scattering amplitudes in reaction channels are given by
\beq{4.11}
f_{ij}({\bf k},{\bf k}')=-\sum_{j',\nu}\phi^{j'}_{\nu}
\av{j|e^{i({\bf k}-{\bf k}')\hat{\bf R}_{\nu }}|j'}.
\eeq
The final momentum ${\bf k}'\equiv{\bf k}_j$ is defined by
${\bf k}_j=k_j({\bf r}/r)$.

It is instructive to note that one obtains in Born approximation
with pseudopotential (\ref{1.1}) the following expression for the
scattering amplitude:
\beq{4.11b}
f_{ij}({\bf k},{\bf k}')=-\sum_{\nu}b_{\nu}
\av{j|e^{i({\bf k}-{\bf k}')\hat{\bf R}_{\nu }}|i}.
\eeq
It is analogous to (\ref{4.11}) but instead of $\phi^j_{\nu }$ it
has the scattering amplitude on an bound nucleus, and the matrix
element is taken between unperturbed states $\langle j|$ and
$|i\rangle$.

{\it Remark}. A typical target model used for ultracold neutron
reflection is semi-infinite substance. The asymptotic procedure
used above to extract the scattering amplitude (\ref{4.11}) from
the wave function, strictly speaking, is incorrect for an infinite
target. So, we shall assume our target finite with some plane
surface area $S$, which is large enough to use in parallel planes
continuum spectrum, orto-normalized with
$\delta({\bf k}'_{\parallel}-{\bf k}_{\parallel})$ instead of
$\delta_{{\bf k}'_{\parallel}{\bf k}_{\parallel}}$, but allows
when necessary to make replacement
\beq{4.12}
\left[(2\pi)^2\delta({\bf k}'_{\parallel}-
{\bf k}_{\parallel})\right]^2
\longrightarrow S(2\pi)^2
\delta({\bf k}'_{\parallel}-{\bf k}_{\parallel}).
\eeq
In the case when transmission is also of importance we need to
consider the target of finite width to allow asymptotic in both
directions.

The cross section for the reaction $i\to j$ can be obtained from
the scattering amplitude by
\beq{4.13}
d\sigma _{ij}=\frac{k'}{k}
|f_{ij}({\bf k},{\bf k}')|^2d\Omega',
\eeq
but the final state of the target is usually not fixed and
inelastic processes are measured by the energy transfer. This
process is described by double differential cross section
\beq{4.13a}
\frac{d^2\sigma}{d\Omega'd\epsilon'}=
\frac{k'}{k}\sum_j|f_{ij}({\bf k},{\bf k}')|^2
\delta(\varepsilon_i-\varepsilon_j+
\epsilon_{{\bf k}}-\epsilon_{{\bf k}'}).
\eeq

It is useful to consider the quantity
\beq{4.14}
w({\bf k},{\bf k}')=\sum_j\left|\frac{2\pi}{m}
f_{ij}({\bf k},{\bf k}')\right|^2
2\pi\delta(\varepsilon_i-\varepsilon_j+
\epsilon_{{\bf k}}-\epsilon_{{\bf k}'}),
\eeq
then $w({\bf k},{\bf k}')d{\bf k}'/(2\pi)^3$ is the scattering
probability per unit time from the fixed state $|{\bf k}\rangle$
to the states $|{\bf k}'\rangle$ into the momentum space element
$d{\bf k}'$. The scattering probability (\ref{4.14}) divided by
the incident neutron flux $k/m$ gives the cross section to find
the final neutron with momentum ${\bf k}'$, and one is free to
choose parameters in ${\bf k}'$ to be fixed in the final state and
then integrate over the rest of these parameters. The relation
between $w({\bf k},{\bf k}')$ and the cross section is evident
from the equality
\beq{4.15}
\frac{m}{k}\int w({\bf k},{\bf k}')\frac{d{\bf k}'}{(2\pi)^3}=
\int\frac{d^2\sigma}{d\Omega'd\epsilon'}
d\Omega'd\epsilon'.
\eeq
So, the cross section of inelastic scattering with energy loss
$\omega$ is given by
\beq{4.16}
\frac{d\sigma}{d\omega}=
\frac{m}{k}\int w({\bf k},{\bf k}')
\delta(\epsilon_{{\bf k}}-\epsilon_{{\bf k}'}-\omega)
\frac{d{\bf k}'}{(2\pi)^3}.
\eeq

Replacing delta-function in (\ref{4.14}) by the integral over $t$
and using (\ref{4.11}) one can sum in (\ref{4.14}) over the final
states $j$ as
\beq{4.17}
\begin{array}{l}
{\ds\sum_{j''}}e^{i(\varepsilon_j-\varepsilon_{j''})t}
\av{j|e^{-i({\bf k}-{\bf k}')\hat{\bf R}_{\nu}}|j''}
\av{j''|e^{i({\bf k}-{\bf k}')\hat{\bf R}_{\nu'}}|j'}={}
\\[\medskipamount]
\phantom{{\ds\sum_{j''}}e^{i(\varepsilon_j-\varepsilon_{j''})t}}=
\chi _{jj'}(\nu\nu',{\bf k}-{\bf k}',t),
\end{array}
\eeq
where $\chi _{jj'}(\nu\nu',\vkappa,t)$ (as well as its Fourier
transform) was introduced by (\ref{5.19f}). Then (\ref{4.14})
takes the form
\beq{4.19}
w({\bf k},{\bf k}')=\left(\frac{2\pi}{m}\right)^2
\sum_{jj'\nu\nu'}
\phi^{j*}_{\nu}\phi^{j'}_{\nu'}
\chi_{jj'}(\nu\nu',{\bf k}-{\bf k}',\omega+
\varepsilon_i-\varepsilon_j).
\eeq
The corresponding equation for double differential cross section
is
\beq{4.21}
\frac{d^2\sigma}{d\Omega d\omega}=
\frac{k'}{2\pi k}\sum_{jj'\nu\nu'}
\phi^{j*}_{\nu}\phi^{j'}_{\nu'}
\chi_{jj'}(\nu\nu',{\bf k}-{\bf k}',\omega+
\varepsilon_i-\varepsilon_j).
\eeq
We have now the scattering amplitude (\ref{4.11}) and scattering
probability (\ref{4.19}) expressed by the neutron amplitudes
$\phi^j_{\nu}$, which are determined by the equation
(\ref{5.19d}).

\section{The case of thermal and cold neutrons}
\label{s6}

Here we address to the neutrons with momenta in the range
\beq{5.20}
\sqrt{4\pi n\alpha}\ll k\ll 1/\alpha,
\eeq
for which the rescattering processes are not important and may be
considered as small perturbation.

In the first approximation the last term in (\ref{5.19d}) may be
neglected and we obtain
\beq{5.21}
\phi^j_{\nu }=\delta_{ij}\frac{\beta_{\nu}}
{1+i\av{i|\hat K_{i\nu}|i}\alpha_{\nu}}\simeq
\delta_{ij}\beta_{\nu }\left(1-i\av{i|\hat
K_{i\nu}|i}\alpha_{\nu}\right),
\eeq
which is similar to the expression (\ref{1.2}) for the amplitude
of neutron scattering on the isolated nucleus. The only difference
is that instead of impact momentum in (\ref{1.2}) (natural
parameter for two-body problem), in (\ref{5.21}) enters the
average value of this parameter over nuclear ensemble in the
target. After substitution of (\ref{5.21}) into (\ref{4.21}) we
obtain the known expression (\ref{1.3}) for thermal neutron
scattering cross section.

\section{Renormalized amplitudes for condensed target}

Up to now no assumptions were made on the target matter. For a
condensed target the results can be presented in more visual form.

Let suppose
\beq{5.22}
{\bf R}_{\nu }=\vrho_{\nu}+{\bf u}_{\nu},
\eeq
where for solid state target $\vrho_{\nu}$ is the equilibrium
position of the target nucleus, and
${\bf u}_{\nu}$ is the shift from the equilibrium. For liquids
$\vrho_{\nu}$ and ${\bf u}_{\nu}$ may be understood as an average
position and a fluctuation.  Note, that index $\nu$ may be now
replaced by $\vrho$. When useful we shall make such replacement
without notice.

The matrix $\chi_{jj'}(\nu\nu',\vkappa,t)$ is transformed now into
\beq{5.23}
\chi_{jj'}(\nu\nu',\vkappa,t)=
e^{-i\vkappa(\vrho_{\nu}-\vrho_{\nu'})}
\av{j|e^{-i\vkappa\hat{\bf u}_{\nu}(t)}
e^{i\vkappa\hat{\bf u}_{\nu'}(0)}|j'}.
\eeq
For the model with fixed (unmovable) nuclei the matrix element in
(\ref{5.23}) equals to $\delta_{jj'}$. In order to separate the
effect of nuclei motion let us write
\beq{5.24}
\chi_{jj'}(\nu\nu',\vkappa,t)=
e^{-i\vkappa(\vrho_{\nu}-\vrho_{\nu'})}
\left(\delta_{jj'}+\tilde\chi_{jj'}(\nu\nu',\vkappa,t)\right),
\eeq
where
\beq{5.25}
\tilde\chi_{jj'}(\nu\nu',\vkappa,t) =
\av{j|e^{-i\vkappa\hat{\bf u}_{\nu }(t)}
e^{i\vkappa\hat{\bf u}_{\nu'}(0)}-1| j'}.
\eeq

To simplify the equation (\ref{5.19d}) let us introduce instead of
$\zeta_{jj'}(\nu\nu')$ (\ref{5.19e}) a new matrix
$G_{jj'}(\nu\nu')$ from
\beq{5.26}
\zeta_{jj'}(\nu\nu')=-e^{-i{\bf k}(\vrho_{\nu}-\vrho_{\nu'})}
G_{jj'}(\nu\nu').
\eeq
Then introducing the new amplitude
\beq{5.27}
\psi_j(\nu)=\frac{1}{\beta_{\nu}}\phi^j_{\nu}
e^{i{\bf k}\vrho_{\nu}},
\eeq
we obtain the equation
\beq{5.28}
\psi_j(\nu)=\delta_{ij}e^{i{\bf k}\vrho_{\nu}}-
\sum_{j'\nu'}G_{jj'}(\nu\nu')\psi_{j'}(\nu')\beta_{\nu'},
\eeq
where the diagonal in $\nu$ and $\nu'$ term is of the form:
$G_{jj'}(\nu\nu)=i(m_{\nu}/m)\av{j|\hat K_{j\nu}|j'}$. Note, that
according to (\ref{5.15}) this diagonal term equals to
$-\zeta_{jj'}(\nu\nu,{\bf x})$ in the limit $x\to 0$ but without
its real part in contrast with the non-diagonal terms. In the long
wave length limit the sum in (\ref{5.28}) may be replaced by the
integral over $\nu'$. In this case it is not necessary to give
special attention to the point $\nu'=\nu$, since the real part of
the term with $\nu'=\nu$ though singular but integrable and does
not contribute into the integral.

For the matrix $G_{jj'}(\nu\nu')$ we have from (\ref{5.19e}),
(\ref{5.24}) and (\ref{5.26})
\beq{5.29}
G_{jj'}(\nu\nu')=\delta_{jj'}G_j(\nu\nu')+\tilde G_{jj'}(\nu\nu')
\eeq
with
\beq{5.30}
G_j(\nu\nu')=-4\pi\sum_{{\bf q}}
\frac{e^{i{\bf q}(\vrho_{\nu}-\vrho_{\nu'})}}
{k_j^2-q^2+i\eta}=
\frac{e^{ik_j|\vrho_{\nu}-\vrho_{\nu'}|}}
{|\vrho_{\nu}-\vrho_{\nu'}|},
\eeq
\beq{5.31}
\tilde G_{jj'}(\nu\nu') =-4\pi \sum_{{\bf q,}\omega}
e^{i{\bf q}(\vrho_{\nu}-\vrho_{\nu'})}
\frac{\tilde\chi_{jj'}(\nu\nu',{\bf k}-{\bf q},\omega)}
{k_j^2-q^2-2m\omega+i\eta},
\eeq
where Fourier transform of $\tilde\chi_{jj'}(\nu\nu',\vkappa,t)$
(\ref{5.25}) is introduced.

The scattering amplitudes in reaction channels (\ref{4.11}) in the
new notations are given by
\beq{5.32}
f_{ij}({\bf k},{\bf k}')=-\sum_{j',\nu}
e^{-i{\bf k}'\vrho_{\nu}}
\av{j|e^{i({\bf k}-{\bf k}')\hat{\bf u}_{\nu}}|j'}
\psi_{j'}(\nu)\beta_{\nu}.
\eeq
The scattering probability can be found with (\ref{5.32}) and
(\ref{4.14})
or directly from (\ref{4.19}) with (\ref{5.27}) and (\ref{5.28}).

{\it Remark.} In Appendix we obtain an alternative form of
(\ref{5.28}) and (\ref{5.32}), which may be useful for some
approximations.

\section{Unitarity condition}

The flux of neutrons incident on the target should be equal to the
total flux in all scattering and reaction channels. As will be
proved in this section, this unitarity condition is satisfied by
the general theory suggested.

It is convenient to rewrite general equation (\ref{5.28}) for
neutron amplitudes in the form
\beq{5b.1}
\delta_{ij}e^{i{\bf k}\vrho_{\nu}}=
\psi_j(\nu)+\sum_{j'\nu'}G_{jj'}(\nu\nu')
\psi_{j'}(\nu')\beta_{\nu'},
\eeq
and after complex conjugation
\beq{5b.2}
\delta_{ij}e^{-i{\bf k}\vrho_{\nu}}=
\psi_j^*(\nu)+\sum_{j'\nu'}G_{jj'}^*(\nu\nu')
\psi_{j'}^*(\nu')\beta^*_{\nu'}.
\eeq
Multiplying (\ref{5b.1}) by $\psi_j^*(\nu)\beta^*_{\nu}$ and
(\ref{5b.2}) by $\psi_j(\nu)\beta_{\nu}$, then summing over $j$,
$\nu$ and subtracting one equation from the other we get
\beq{5b.3}
\begin{array}{l}
2i\,{\rm Im}f_{ii}({\bf k},{\bf k})=
{\ds\sum_{j\nu}}
(-2i\,{\rm Im}\beta_{\nu})\psi_j^*(\nu)\psi_j(\nu)+{}
\\[\medskipamount]
\phantom{}+
{\ds\sum_{jj'\nu\nu'}}
\psi_j^*(\nu)\psi_{j'}(\nu')\beta^*_{\nu}\beta_{\nu'}
\left(G_{jj'}(\nu\nu')-G_{j'j}^*(\nu'\nu)\right),
\end{array}
\eeq
where we take into account equation (\ref{5.32}) for scattering
amplitude.

The kernels $G$ and $G^*$, as seen from (\ref{5.19e}) and
(\ref{5.26}), differ only in path directions around the poles on
complex $q$ plane, and their difference can be presented in the
form
\beq{5b.4}
\begin{array}{l}
G_{jj'}(\nu\nu')-G_{j'j}^*(\nu'\nu)=
-\Frac{2\pi}{m}{\ds\sum_{{\bf q}}}
e^{i{\bf q}(\vrho_{\nu}-\vrho_{\nu'})}\times{}
\\[\medskipamount]
\phantom{}\times
\av{j|e^{-i({\bf k}-{\bf q})\hat{\bf u}_{\nu}}
\left(\hat D^{-1}_q-(\hat D^{-1}_q)^+\right)
e^{i({\bf k}-{\bf q})\hat{\bf u}_{\nu'}}|j'}.
\end{array}
\eeq
Operator $\hat D^{-1}_q$ has the form $(a+i\eta)^{-1}$ and due to
the equality $(a+i\eta)^{-1}=-i\pi\delta(a)+{\rm P}(1/a)$ only
delta-function part remains in combination
$\hat D^{-1}_q-(\hat D^{-1}_q)^+$. Thus (\ref{5b.4}) reduces to
\beq{5b.5}
\begin{array}{l}
G_{jj'}(\nu\nu')-G_{j'j}^*(\nu'\nu)=
\Frac{i}{2\pi}{\ds\int}d{\bf n}''{\ds\sum_{j''}}k_{j''}
e^{i{\bf k}_{j''}(\vrho_{\nu}-\vrho_{\nu'})}\times{}
\\[\medskipamount]
\phantom{G_{jj'}(\nu\nu')}\times
\av{j|e^{-i({\bf k}-{\bf k}_{j''})\hat{\bf u}_{\nu}}|j''}
\av{j''|e^{i({\bf k}-{\bf k}_{j''})\hat{\bf u}_{\nu'}}|j'},
\end{array}
\eeq
where ${\bf n}''={\bf k}_{j''}/k_{j''}$.

Substituting this result in (\ref{5b.3}) and taking into account
(\ref{5.32}) we get total cross section $\sigma_t$ as sum of
capture cross section $\sigma_c$ and total scattering cross
section $\sigma_s$:
\beq{5b.6}
\sigma_t=\frac{4\pi}{k}{\rm Im}f_{ii}({\bf k},{\bf k})=
\sigma_c+\sigma_s,
\eeq
where
\beq{5b.7}
\sigma_c=\sum_{j\nu}
(-\frac{4\pi{\rm Im}\beta_{\nu}}{k})
|\psi_j(\nu)|^2,
\eeq
\beq{5b.8}
\sigma_s=\sum_j\frac{k_j}{k}
\int |f_{ij}({\bf k},{\bf k}')|^2d{\bf n}'.
\eeq
Note that capture cross section is determined by imaginary parts
of scattering lengths $\alpha_{\nu}$ and
$\beta_{\nu}=m\alpha_{\nu}/m_{\nu}$ ($-4\pi{\rm Im}\alpha/k$ is
the capture cross section for free nucleus).

\section{Interim summary}

Let us sum up our results. To satisfy MST equations one has to
find N state vectors $\Psi_{\nu}$ and N operators $\hat t_{\nu}$.
We have reduced the problem to the set of linear equations
(\ref{5.19d}) for numerical parameters $\phi^j_{\nu}$. The price
for this reduction is additional index $j$ which arises due to
introduction of unknown state vector $|\nu\rangle$ for each target
nucleus. However, if $\phi^j_{\nu}=\av{j|\nu}\phi_{\nu}$ are found
then one can deduce $\Psi_{\nu}$ and $\hat t_{\nu}$.

Indeed, equations (\ref{5.17}) and (\ref{5.19d}) were directly
obtained from (\ref{5.1}), where the left-hand side has the
meaning of $\av{{\bf R}_{\nu}|\Psi_{\nu}}$. Following back from
(\ref{5.17}) to (\ref{5.1}) one can deduce for $\Psi_{\nu}$
\beq{6.0a}
|\Psi_{\nu}\rangle=|{\bf R}_{\nu}\rangle
e^{i{\bf k}\hat{\bf R}_{\nu}}\frac{1}{\beta_{\nu}}
\left(1+i\alpha_{\nu}\hat K_{\nu}\right)
|\nu\rangle\phi_{\nu},
\eeq
where
\beq{6.0b}
\hat K_{\nu}=\sum_j|j\rangle\langle j|\hat K_{j\nu}.
\eeq
On the other hand, if to compare (\ref{5.19g}) with (\ref{1.10}),
one can obtain combination
\beq{6.0c}
\hat t_{\nu}|\Psi_{\nu}\rangle=
\frac{2\pi}{m}|{\bf R}_{\nu}\rangle
e^{i{\bf k}\hat{\bf R}_{\nu}}|\nu\rangle\phi_{\nu}.
\eeq
From (\ref{6.0b}) and (\ref{6.0c}) it follows for $\hat t_{\nu}$
\beq{6.0d}
\hat t_{\nu}=\frac{2\pi}{m}
|{\bf R}_{\nu}\rangle e^{i{\bf k}\hat{\bf R}_{\nu}}
\left(1+i\alpha_{\nu}\hat K_{\nu}\right)^{-1}
\beta_{\nu}\,e^{-i{\bf k}\hat{\bf R}_{\nu}}\langle{\bf R}_{\nu}|.
\eeq
This expression for $\hat t_{\nu}$ is rather complicated due to
non-commuting in target space operators $\hat K_{\nu}$ and
$\hat {\bf R}_{\nu}$.

{\it Remark.} In (\ref{2.3b}) $\Psi_{\nu}$ seems to be a state
vector in N+1 dimensional space
$({\bf r},{\bf R})$. However, for short-range $\hat t_{\nu}$ the
total wave function $\Psi_{{\bf k},i}$ is really determined by
$\Psi_{\nu}$ at ${\bf r}={\bf R}_{\nu}$. This part, in fact, is
given by (\ref{6.0a}).

For fixed nuclei $\hat K_{\nu}\to k_{\nu}$ is the center-of-mass
impact momentum and from (\ref{6.0d}) follows
\beq{6.0e}
\av{{\bf r}|\hat t_{\nu}|{\bf r}'}\longrightarrow
\frac{2\pi}{m}\,\frac{\beta_{\nu}}{1+i\alpha_{\nu}k_{\nu}}\,
\delta({\bf r}-{\bf R}_{\nu})
\delta({\bf r}-{\bf r}'),
\eeq
that, with the help of (\ref{1.2}) and (\ref{5.18}), coincides
with Fermi pseudopotential (\ref{1.1}). Just in this approximation
(\ref{1.8}) and (\ref{1.9}) follow from (\ref{1.10}) and
(\ref{1.11}), respectively.

Fortunately, in general case all physical quantities can be
obtained directly from the amplitudes $\phi^j_{\nu}$ or
$\psi_j(\nu)$ and matrices $\hat t_{\nu}$ are not needed. Thus we
reduce MST equations (\ref{1.10}) and (\ref{1.11}) to systems
(\ref{5.19d}) for $\phi^j_{\nu}$ or (\ref{5.28}) for
$\psi_j(\nu)$.

Let us emphasize, that the only approximations used so far are
those for the interaction potential. It was assumed a short range
and relatively deep, what is equivalent to scattering length
approximation for the interaction.

A relation of our equations to the traditional theory for slow
neutrons was demonstrated in section~\ref{s6}. For thermal and
cold neutrons the structure of equations (\ref{5.19d}) may be
radically simplified due to the physically justifiable neglect of
rescattering of the secondary neutron waves in the target.

Our main goal here is UCN. A relation to the traditional theory
for the (elastic) scattering of UCN will be in details considered
below. Physically evident, that transition to that theory should
occur by the neglect of the thermal motion of the target nuclei,
which provides the radical simplification of the correlation
function (\ref{5.23}):
\beq{6.1}
\chi_{jj'}(\nu\nu',\vkappa,t)\simeq
\delta_{jj'}e^{-i\vkappa(\vrho_{\nu}-\vrho_{\nu'})}.
\eeq

The amplitudes of the thermal motion are indeed small as compared
to the UCN wave length and the approximation (\ref{6.1}) for
$\chi_{jj'}$ seams to be justified. But if the thermal motion is
totally neglected it is no possibility for inelastic scattering.
So, for inelastic processes we need
to include the thermal motion but may hope for proper
simplification since a perturbation procedure is justified.

The rest of the paper is devoted to a perturbational solution of
the main equation (\ref{5.28}) for $\psi_j(\nu)$.

\section{Expansion over the amplitudes of thermal vibrations}

\subsection{Zero order (elastic) approximation}

In the long wave length limit ($\vkappa {\bf u}\ll 1$) it follows
from (\ref{5.25})
\beq{7.1}
\tilde\chi_{jj'}(\nu\nu',\vkappa,t)\ll 1,
\eeq
which in its turn leads to the inequality
\beq{7.2}
\tilde G_{jj'}(\nu\nu')\ll G_j(\nu\nu').
\eeq
These inequalities open a possibility for the application of a
perturbation theory.

Physically (\ref{7.1}) and (\ref{7.2}) mean that in this limit the
basic process of neutron -- target interaction is elastic
scattering and the probability of inelastic processes is small.

If we neglect $\tilde\chi_{jj'}$ then we obtain from (\ref{5.28})
an equation for the amplitude $\psi_j^{(0)}(\nu)$ in zero order
approximation
\beq{7.3}
\psi_j^{(0)}(\nu)=\delta_{ij}e^{i{\bf k}\vrho_{\nu}}-
\sum_{\nu'}G_j(\nu\nu')\psi_j^{(0)}(\nu')\beta_{\nu'}.
\eeq
As seen from (\ref{7.3}), the equations for different $j$ are
separated, and since inhomogeneous term contains the factor
$\delta_{ij}$, we have
\beq{7.4}
\psi_j^{(0)}(\nu)=\delta_{ij}\psi(\nu),
\eeq
where $\psi(\nu)$ is defined by the equation
\beq{7.5}
\psi(\nu)=e^{i{\bf k}\vrho_{\nu }}-
\sum_{\nu'}G(\nu\nu')\psi(\nu')\beta_{\nu'}
\eeq
with the matrix
\beq{7.6}
G(\nu\nu')\equiv G_i(\nu\nu') =
-4\pi\sum_{{\bf q}}\frac{e^{i{\bf q}(\vrho_{\nu}-\vrho_{\nu'})}}
{k^2-q^2+i\eta}=
\frac{e^{ik|\vrho_{\nu}-\vrho_{\nu'}|}}
{|\vrho_{\nu}-\vrho_{\nu'}|}.
\eeq
Note, that singular real part of diagonal ($\nu'=\nu$) term in
(\ref{7.6}) as discussed after (\ref{5.28}) should be extracted
that results in $G(\nu\nu)=ik$.

The equation (\ref{7.5}) is similar to formula (\ref{1.9})
customary used for UCN in the target model with fixed (i.e. in
fact, infinitely heavy) nuclei and transforms to it by
redefinition
$\Psi_{{\bf k}}(\nu)=\psi(\nu)(1+ik\beta_{\nu})$ \cite{Bar99}. Its
solution is basically simplified when the sum over $\nu$ is
replaced by the integral over $\vrho$. Then after acting on
(\ref{7.5}) with operator $\Delta+k^2$ it is reduced to
Schr\"{o}dinger equation with the potential
\beq{7.7}
U=\frac{2\pi}{m}n\beta,
\eeq
where the density of the target $n$ and scattering length $\beta$
may depend on $\vrho$.

Formally, the equation (\ref{7.5}) after replacement of the sum by
integral
$\sum_{\nu}\to\int nd\vrho$, becomes of a linear integral type.
Such a replacement, if useful, will be performed below without
special notice. On the other hand, to make presentation of general
formulae in more compact and transparent form it is useful to
consider all $G(\nu\nu')$ as matrices, $\psi(\nu)$ and
$e^{i{\bf k}\vrho_{\nu}}$ -- as columns, $\psi^*(\nu)$ and
$e^{-i{\bf k}\vrho_{\nu}}$ -- as rows and omit summation indices
$\nu$ and $\nu'$. In this notation (\ref{7.5}) looks as
\beq{7.8}
\psi+G\psi\beta=e^{i{\bf k}\vrho}.
\eeq
As seen from (\ref{7.6}), the kernel $G(k,\nu\nu')$ depends on the
absolute value $k$, but a solution $\psi({\bf k},\nu)$ depends on
the vector ${\bf k}$, defined by inhomogeneous term on the
right-hand side.

The scattering amplitude (\ref{5.32}) in zero order approximation
is given by
\beq{7.9}
f^{(0)}_{ij}({\bf k},{\bf k}')=-\delta_{ij}\psi({\bf k},{\bf k}'),
\eeq
where special notation is introduced for $\beta$-weighted
Fou\-rier-transform of the amplitude $\psi({\bf k},\nu)$
\beq{7.10}
\psi({\bf k},{\bf q})=\sum_{\nu}
e^{-i{\bf q}\vrho_{\nu }}\psi({\bf k},\nu)\beta_{\nu}.
\eeq
Scattering probability (\ref{4.14}) in zero order approximation is
of the form
\beq{7.11}
w^{(0)}({\bf k},{\bf k}')=\frac{(2\pi)^3}{m^2}
\delta(\omega)\left|\psi({\bf k},{\bf k}')\right|^2.
\eeq
Thus from (\ref{4.15}) we obtain differential cross section of
elastic scattering
\beq{7.12}
\frac{d\sigma^{(0)}_{el}}{d\Omega}=
\left|\psi({\bf k},{\bf k}')\right|^2,
\eeq
where $|{\bf k}'|=|{\bf k}|$.

\subsection{Approximations of the first and the second order.
Inelastic scattering}

Zero order amplitude given by (\ref{7.9}) corresponds to elastic
scattering. Thus, the probability (\ref{4.14}) for inelastic
scattering starts from the second order term which is determined
by the first order scattering amplitude
\beq{7.13}
w_{ie}^{(2)}({\bf k},{\bf k}')=
2\pi\sum_j\left|\frac{2\pi}{m}f^{(1)}_{ij}({\bf k},
{\bf k}')\right|^2
\delta(\varepsilon_i-\varepsilon_j+\epsilon_{{\bf k}}-
\epsilon_{{\bf k}'}).
\eeq
Introducing an expansion in $\vkappa{\bf u}$ for neutron
amplitudes
\beq{7.14}
\psi_j(\nu)=\delta_{ij}\psi(\nu)+
\psi^{(1)}_j(\nu)+\ldots,
\eeq
one obtains from (\ref{5.32})
\beq{7.15}
\begin{array}{l}
f^{(1)}_{ij}({\bf k},{\bf k}')=
-\psi^{(1)}_j({\bf k},{\bf k}')-{}
\\[\medskipamount]
\phantom{}-
i(k^{\sigma}-k'^{\sigma}){\ds\sum_{\nu}}
e^{-i{\bf k}'\vrho_{\nu}}
\av{j|\hat u_{\nu}^{\sigma}|i}\psi(\nu)\beta_{\nu},
\end{array}
\eeq
where the first term is $\beta$-weighted Fourier component
(\ref{7.10}) of the first order amplitudes $\psi^{(1)}_j$ defined
by the equation
\beq{7.16}
\psi^{(1)}_j+G_j\psi^{(1)}_j\beta=-\tilde G^{(1)}_{ji}\psi\beta.
\eeq
Matrix $\tilde G^{(1)}_{jj'}(\nu\nu')$ is the first order term
which follows from (\ref{5.25}) and (\ref{5.31})
\beq{7.17}
\begin{array}{l}
\tilde G^{(1)}_{jj'}(\nu\nu')=
\left(\nabla_{\nu}^{\sigma}-ik^{\sigma}\right)
\Bigl(\av{j|\hat u_{\nu}^{\sigma}|j'}G_{j'}(\nu\nu')-{}
\\[\medskipamount]
\phantom{\tilde G_{jj'}(\nu\nu')=
\left(\nabla_{\nu}^{\sigma}-ik^{\sigma}\right)}-
G_j(\nu\nu')\av{j|\hat u_{\nu'}^{\sigma}|j'}\Bigr),
\end{array}
\eeq
where
$\nabla_{\nu}^{\sigma}=\partial/\partial\rho_{\nu}^{\sigma}$.

Fourier component $\psi^{(1)}_j({\bf k},{\bf k}')$ can be found
without an explicit solution of equation (\ref{7.16}). First, let
us introduce the function
\beq{7.18}
\bar\psi({\bf k}',\nu)=\psi(-{\bf k}',\nu),
\eeq
determined by the equation
\beq{7.19}
\bar\psi+\beta\bar\psi\bar G=e^{-i{\bf k}'\vrho},
\eeq
where we consider $\bar\psi$ as row, and $\bar G(\nu\nu')\equiv
G_j(\nu\nu')$. Then, multiplying (\ref{7.16}) by $\beta\bar\psi$
from the left and using (\ref{7.19}), we obtain in the left-hand
side just the Fourier component that we are looking for. Therefore
we have
\beq{7.20}
\psi^{(1)}_j({\bf k},{\bf k}')=
-\beta\bar\psi\tilde G^{(1)}_{ji}\psi\beta.
\eeq

Right-hand side of this equation can be simplified with the help
of (\ref{7.8}), (\ref{7.19}) and (\ref{7.17}). Then, we finally
obtain
\beq{7.21}
\begin{array}{l}
\psi^{(1)}_j({\bf k},{\bf k}')=
{\ds\sum_{\nu}}\beta_{\nu}\av{j|\hat u_{\nu}^{\sigma}|i}
\nabla_{\nu}^{\sigma}\left(\bar\psi(\nu)\psi(\nu)\right)-{}
\\[\medskipamount]
\phantom{}-
i(k^{\sigma}-k'^{\sigma})
{\ds\sum_{\nu}}e^{-i{\bf k}'\vrho_{\nu}}
\av{j|\hat u_{\nu}^{\sigma}|i}\psi(\nu)\beta_{\nu}.
\end{array}
\eeq
Inserting (\ref{7.21}) into (\ref{7.15}) we get for the first
order scattering amplitude
\beq{7.22}
f^{(1)}_{ij}({\bf k},{\bf k}')=
-\sum_{\nu}\beta_{\nu}\av{j|\hat{\bf u}_{\nu}|i}
\mbox{\boldmath $\nabla$}_{\nu}
\left(\bar\psi({\bf k}',\nu)\psi({\bf k},\nu)\right).
\eeq

{\it Remark.} An alternative derivation of expressions (\ref{7.9})
and (\ref{7.22}) is given in Appendix.

The probability of inelastic scattering (in the second order)
follows from (\ref{7.13}) and (\ref{7.22})
\beq{7.23}
\begin{array}{l}
w_{ie}^{(2)}({\bf k},{\bf k}')=
\left(\Frac{2\pi}{m}\right)^2{\ds\sum_{j\nu\nu'}}
\beta^*_{\nu}\beta_{\nu'}\times{}
\\[\bigskipamount]
\phantom{}\times
\nabla_{\nu}^{\sigma}
\left(\bar\psi({\bf k}',\nu)\psi({\bf k},\nu)\right)^*
\nabla_{\nu'}^{\tau}
\left(\bar\psi({\bf k}',\nu')\psi({\bf k},\nu')\right)\times{}
\\[\bigskipamount]
\phantom{}\times
{\ds\int\limits_{-\infty}^{+\infty}}
e^{i(\varepsilon_i-\varepsilon_j+\epsilon_{{\bf k}}-
\epsilon_{{\bf k}'})t}\,
\av{i|\hat u_{\nu}^{\sigma}|j}
\av{j|\hat u_{\nu'}^{\tau}|i}\,dt.
\end{array}
\eeq
Summation over $j$ can be explicitly performed according to
\beq{7.24}
\sum_je^{i(\varepsilon_i-\varepsilon_j)t}
\av{i|\hat u_{\nu}^{\sigma}|j}
\av{j|\hat u_{\nu'}^{\tau}|i}=
\av{i|\hat u_{\nu}^{\sigma}(t)\hat u_{\nu'}^{\tau}(0)|i}.
\eeq
This diagonal matrix element may exhibit a spatial dependence only
as a function of $\vrho_{\nu}-\vrho_{\nu'}$ and therefore allows a
Fourier transform
\beq{7.25}
\av{i|\hat u_{\nu}^{\sigma}(t)\hat u_{\nu'}^{\tau}(0)|i}=
\sum_{{\bf q},\omega}e^{i{\bf q}(\vrho_{\nu}-\vrho_{\nu'})-
i\omega t}\,
\Omega^{\sigma\tau}({\bf q},\omega).
\eeq

Then, we finally obtain for the probability of inelastic
scattering
\beq{7.26}
\begin{array}{l}
w_{ie}^{(2)}({\bf k},{\bf k}')={}
\\[\medskipamount]
\phantom{}=
\Frac{(2\pi)^3}{m^2}
{\ds\sum_{{\bf q},\omega}}\delta(\epsilon_{{\bf k}}-
\epsilon_{{\bf k}'}-\omega)
B^{\sigma *}({\bf q})B^{\tau}({\bf q})
\Omega^{\sigma\tau}({\bf q},\omega),
\end{array}
\eeq
where
\beq{7.27}
{\bf B}({\bf q})=
\sum_{\nu}\beta_{\nu}e^{-i{\bf q}\vrho_{\nu}}
\mbox{\boldmath $\nabla$}_{\nu}
\left(\bar\psi({\bf k}',\nu)\psi({\bf k},\nu)\right).
\eeq
The cross section for neutron to lose energy $\omega$ can be
calculated from (\ref{4.16}) and (\ref{7.26})
\beq{7.28}
\frac{d\sigma_{ie}^{(2)}}{d\omega}=
\frac{(2\pi)^2}{mk}\sum_{{\bf q},{\bf k}'}
\delta(\epsilon_{{\bf k}}-\epsilon_{{\bf k}'}-\omega)
B^{\sigma *}({\bf q})
B^{\tau}({\bf q})\Omega^{\sigma\tau}({\bf q},\omega).
\eeq

To disclose physical meaning of (\ref{7.28}) it is instructive to
compare it with corresponding expression which followed from
(\ref{1.3}). To make the comparison one should transform the
correlation function $\chi(\nu\nu',\vkappa,\omega)$ to
$\Omega^{\sigma\tau}({\bf q},\omega)$ (\ref{7.25}). Expanding
$\chi(\nu\nu',\vkappa,\omega)$ in $\vkappa{\bf u}$  by the use of
(\ref{5.22}) and (\ref{5.23}) we get
\beq{7.29}
\begin{array}{l}
\chi(\nu\nu',\vkappa,\omega)=
2\pi\delta(\omega)e^{-i\vkappa(\vrho_{\nu}-\vrho_{\nu'})}
\left(1-\av{(\vkappa\hat{\bf u}_{\nu})^2}\right)+{}
\\[\medskipamount]
\phantom{\chi(\nu\nu')}+
e^{-i\vkappa(\vrho_{\nu}-\vrho_{\nu'})}
\kappa^{\sigma}\kappa^{\tau}
{\ds\int\limits_{-\infty}^{+\infty}}
\av{i|\hat u_{\nu}^{\sigma}(t)\hat u_{\nu'}^{\tau}(0)|i}
e^{i\omega t}dt.
\end{array}
\eeq

Now it is easy to see that cross section with energy loss (see
(\ref{4.15}) and (\ref{4.16})), followed from (\ref{1.3}) and
(\ref{7.29}), can be reduced to the form (\ref{7.28}) with
\beq{7.30}
\tilde{\bf B}({\bf q})=i({\bf k}-{\bf k}')
\sum_{\nu}b_{\nu}e^{-i{\bf q}\vrho_{\nu}}
e^{i({\bf k}-{\bf k}')\vrho_{\nu}}
\eeq
instead of ${\bf B}({\bf q})$ (\ref{7.27}). The difference (apart
from factor N) is that functions $\psi({\bf k},\nu)$ and
$\bar\psi({\bf k}',\nu)$ in (\ref{7.27}) are replaced by the plane
waves $e^{i{\bf k}\vrho_{\nu}}$ and $e^{-i{\bf k}'\vrho_{\nu}}$,
respectively. It is very natural since (\ref{1.3}) is obtained in
Born approximation.

So, one may say that (\ref{7.28}), similar to (\ref{1.3}), takes
into account interference of two scattered waves (that result in
inelasticity), but in addition uses wave functions in both input
and output channels modified by rescattering.

The idea to modify (\ref{1.3}) for UCN by replacing plane waves
with solutions of equation (\ref{1.9}) is very natural and was
tried in several papers (see, e.g., \cite{Blo77}). But it is
evident that if to do it in (\ref{7.30}) the result will not
coincide with (\ref{7.27}).

\section{Choice of correlation function}

Inelastic processes with energy and momentum exchange between
neutron and target are essentially determined by the dynamical
properties of target matter, i.e. collective excitations that are
suitable (for given conservation lows) to provide this exchange.
Correlation function that enters into cross section just describes
these dynamical properties. The physical meaning of correlation
functions -- description of space--time evolution of a fluctuation
appeared at some moment in some position point.

The field of correlation functions is covered in a number of books
and review articles (see, e.g., \cite{Lov84,For75,Han86}). So,
here we'll just mention a few details necessary for what follows.

Our function (\ref{7.25}) is related to density fluctuations
\beq{8.1}
\frac{1}{2\pi N}\av{\int
\hat n({\bf r}'+{\bf r},t)\,\hat n({\bf r}',0)\,d{\bf r}'\,}=
\sum_{{\bf q},\omega}e^{i{\bf q}{\bf r}-i\omega t}
S({\bf q},\omega).
\eeq
Fourier transform $S({\bf q},\omega)$ (often denoted as
''dynamical structure factor'') can be shown to be connected with
(\ref{7.25}) by
\beq{8.2}
S({\bf q}',\omega)\simeq\frac{n}{2\pi}
q^{\sigma}q^{\tau}\Omega^{\sigma\tau}({\bf q},\omega) ,
\eeq
where the quantities ${\bf q}'$ and ${\bf q}$ are equal but for
crystals may differ by a reciprocal lattice vector.

For simple model of harmonic crystal one can easily obtain (for
phonon occupation factors
$n_{{\bf q}}\gg 1$)
\beq{8.3}
\Omega^{\sigma\tau}({\bf q},\omega)\simeq
\delta_{\sigma\tau}\,\frac{2T}{nMs^2}\,
\frac{\pi}{|\omega|}\,
\delta\left(q^2-\frac{\omega^2}{s^2}\right),
\eeq
where $T$ is temperature and $s$ is velocity of sound.

Sound branch of excitation is effective for large energy and
momentum transfer (say, from UCN to thermal), but very ineffective
for small transfer (of the order of initial energy and momentum of
UCN). The reason is in fact that neutron dispersion low
$\epsilon\sim vk$ is quite different from that of sound
$\omega=sq$ because for UCN $v/s\sim 10^{-3}$ and one cannot
satisfy two requirements $\Delta\epsilon\sim\omega$ and $\Delta
k\sim q$ simultaneously. For small transfers we need excitations
with small $\omega$ and ${\bf q}$.

Limiting value of correlation function for $\omega, q\to 0$ is
given by ''hydrodynamic value''
\beq{8.4}
\Omega^{\sigma\tau}({\bf q},\omega)\simeq
\frac{q^{\sigma}q^{\tau }}{q^2}\,
\frac{2T}{nMs^2}\,
\frac{\alpha D}{\omega^2+D^2q^4},
\eeq
where $D$ is coefficient of any diffusion-like process, e.g.,
self- or thermo-diffusion coefficient (in the last case
$\alpha=c_P/c_V-1$, in the first case $\alpha=1$). At normal
temperature parameter $\alpha=c_P/c_V-1$ is of the scale of
10$^{-2}$ for solids and of 10$^{-1}$ for liquids.

Function (\ref{8.4}) for fixed $q$ has a pick value for $\omega=0$
and width
$\sim Dq^2$ in contrast to (\ref{8.3}), where $\omega$ and $q$ are
strongly coupled ($\omega =sq$). It is useful to introduce instead
$D$ a dimentionless parameter
\beq{8.4b}
d=\frac{2mD}{\hbar},
\eeq
which appears if one considers dimentionless variables
$\omega/\epsilon_{{\bf k}}$ and $q/k$. One may expect that the
optimal conditions for small energy $\epsilon\sim\hbar\omega$ and
momentum $k\sim q$ transfer would be when this parameter $d$ is of
the scale of unity. In reality at normal temperature $d$ varies
from $\sim 10^3$ (metals with high thermoconductivity) to
$\sim 10^{-2}$ (self-diffusion in liquids).

The total correlation function include the phonon part (\ref{8.3})
as well as all types of diffusion-like parts (\ref{8.4}). All
these parts are linearly summed in cross section and their
contributions may be calculated separately.

\section{Subbarier inelastic scattering}
\label{s9}

To consider a specific inelastic scattering problem with general
formula (\ref{7.28}) one needs, first, to find zero order
(elastic) neutron amplitudes for input and output channels
$\psi({\bf k},\nu)$ and $\bar\psi({\bf k}',\nu)$ and, second, to
choose a correlation function that is adequately describes
collective excitations of target matter in energy--momentum domain
of interest.

We consider, as illustrative example, scattering on thick uniform
plane target when neutron energies in input and output channels
are both below potential barrier. Let $z$ axis is perpendicular to
the surface of the target located at $z>0$. After replacing
discreet variable $\nu$ by uniform $\vrho$, one can reduce
integral equations (\ref{7.8}) and (\ref{7.19}) to Schr\"{o}dinger
equation
\beq{9.1}
(k^2+\Delta)\psi({\bf r})=u(z)\psi({\bf r}),
\eeq
where potential $u(z)=4\pi\beta n(z)$ is determined by the target
density
\beq{9.2}
n(z)=\left\{\begin{array}{ll}
n, &\quad z>0,\\
0, &\quad z<0.
\end{array}\right.
\eeq
Equations for the cross section contain the values of
$\psi({\bf r})$ and $\bar\psi({\bf r})$ only inside the target,
but solutions for them are determined by the inhomogeneous terms
of integral equations (\ref{7.8}) and (\ref{7.19}) which for
solutions of Schr\"{o}dinger equation (\ref{9.1}) have the meaning
of waves incident on the target. We expand neutron momentum
${\bf k}$ in input channel into components ${\bf k}_{\|}$ and
$k_{\perp}{\bf e}_z$ along and normal to the target surface. We
use in what follows that $k_{\perp}^2\leq k^2<u_0=4\pi\beta n$.

The solution of (\ref{9.1}) for incident neutron in the region
$z>0$ is of the form
\beq{9.3}
\psi({\bf r})=te^{i{\bf k}_{\|}{\bf r}_{\|}-\kps z},\quad
t=\frac{2k_{\perp}}{k_{\perp}+i\kp},\quad
\kp=\sqrt{u_0-k_{\perp}^2}.
\eeq
Neutron momentum in output channel we also expand in sum of
longitudinal and transverse components
${\bf k}'={\bf k}'_{\|}+k'_{\perp}{\bf e}_z$. Thus for the elastic
scattering amplitude (\ref{7.9}) we get
\beq{9.4}
f^{(0)}({\bf k},{\bf k}')=
-\psi({\bf k},{\bf k}')=
2\pi ik_{\perp}\delta^{(2)}({\bf k}'_{\|}-{\bf k}_{\|})
\frac{k'_{\perp}+i\kp}{k_{\perp}+i\kp},
\eeq
where $k'_{\perp}=k_{\perp}$ for transmission and
$k'_{\perp}=-k_{\perp}$ for reflection. Note, that here we take
into account diffraction forward scattering
(${\bf k}'={\bf k}$) which results from finite transverse size of
a target (see remark related to equation (\ref{4.12})).

Substituting (\ref{9.4}) into (\ref{7.12}) and integrating over
solid angle around the direction
${\bf k}'={\bf k}_{\|}-k_{\perp}{\bf e}_z$, we obtain zero order
cross section of neutron elastic reflection from the semi-infinite
target
\beq{9.5}
\sigma^{(0)}_R=S\frac{k_{\perp}}{k}.
\eeq
Here $S$ is area of plain target surface, and
$k_{\perp}/k=\cos\theta$, where $\theta$ is the angle of
incidence. That is simply the whole target area seen from incident
neutron direction. This result is natural for total reflection.

To calculate inelastic scattering to neutron state with momentum
${\bf k}'$ we need the solution
$\bar\psi({\bf k}',{\bf r})=\psi(-{\bf k}',{\bf r})$. Since
subbarier neutron in output channel is back scattered it is
convenient to assume that
${\bf k}'={\bf k}'_{\|}-k'_{\perp}{\bf e}_z$, where
$k'_{\perp}>0$. Thus
\beq{9.6}
\bar\psi({\bf r})=t'e^{-i{\bf k}_{\|}{\bf r}_{\|}-\kps' z},\quad
t'=\frac{2k'_{\perp}}{k'_{\perp}+i\kp'},\quad
\kp'=\sqrt{u_0-k^{\prime 2}_{\perp}}.
\eeq
Vector ${\bf B}$ (\ref{7.27}) is given by
\beq{9.7}
{\bf B}({\bf q})=n\beta
(2\pi)^2\delta^{(2)}({\bf k}_{\|}-{\bf k}'_{\|}-{\bf q}_{\|})
tt'\frac{{\bf q}_{\|}+i(\kp+\kp'){\bf e}_z}
{q_{\perp}-i(\kp+\kp')}.
\eeq

For simplicity we neglect imaginary parts of amplitude $\beta$ and
potential $u_0$ related to radiative capture. Thus, when
substituting (\ref{9.7}) into (\ref{7.28}) we will use
\beq{9.8}
t^*t=\frac{4k^2_{\perp}}{k^2_{\perp}+\kp^2}=
\frac{k^2_{\perp}}{\pi n\beta},
\eeq
and the same for $t^{\prime *}t'$. Taking into account
(\ref{4.12}), one gets for inelastic cross section the following
expression
\beq{9.9}
\begin{array}{l}
\Frac{d\sigma^{(2)}_{ie}}{d\omega}=
S\Frac{k_{\perp}}{k}\,
\Frac{T}{nMs^2}\,
\Frac{k_{\perp}}{\pi^4|\omega|}\,
{\ds\int}d{\bf k}'\,\delta(k'^2+2m\omega-k^2)\times{}
\\[\bigskipamount]
\phantom{\Frac{d\sigma^{(2)}_{ie}}{d\omega}}\times
k^{\prime 2}_{\perp}
{\ds\int}d{\bf q}\,\delta^{(2)}({\bf k}_{\|}-{\bf k}'_{\|}-
{\bf q}_{\|})\,
\Lambda(k'_{\perp},{\bf q}).
\end{array}
\eeq
Here
\beq{9.9b}
\Lambda^{ph}(k'_{\perp},{\bf q})=
\pi\delta\left(q^2-\frac{\omega^2}{s^2}\right)\,
\frac{q^2_{\|}+(\kp+\kp')^2}{q^2_{\perp}+(\kp+\kp')^2}
\eeq
for phonon correlation function (\ref{8.3}) and
\beq{9.9c}
\Lambda^{hyd}(k'_{\perp},{\bf q})=
\Frac{\alpha\Gamma^2}{q^4+\Gamma^4}
\left(\frac{q^2_{\|}+(\kp+\kp')^2}{q^2_{\perp}+(\kp+\kp')^2}-
\frac{q_{\|}^2}{q^2}\right)
\eeq
for hydrodynamic one (\ref{8.4}), where
$\Gamma=\sqrt{|\omega|/D}$. Dividing the inelastic cross section
over the transverse target area $Sk_{\perp}/k$, we get
differential probability per one bounce $dw_{ie}/d\omega$ for
neutron transition to the state with the energy
$\epsilon'=\epsilon-\omega$.

Integration in (\ref{9.9}) over ${\bf q}_{\|}$ removes
two-dimensional delta-function. Then remaining delta-function
allows to perform integration over $k'_{\|}$. The result is
\beq{9.10}
\begin{array}{l}
\Frac{dw_{ie}}{d\omega}=
\Frac{T}{nMs^2}\,
\Frac{k_{\perp}}{\pi^4|\omega|}\,
{\ds\int\limits_0^{\sqrt{k^2-2m\omega}}}k'^2_{\perp}
dk'_{\perp}\times{}
\\[\bigskipamount]
\phantom{\Frac{dw_{ie}}{d\omega}}\times
{\ds\int\limits_0^{\pi}}d\varphi
{\ds\int\limits_{-\infty}^{+\infty}}dq_{\perp}
\Lambda(k'_{\perp},{\bf q}),
\end{array}
\eeq
where one should keep in mind relations
\beq{9.10b}
q^2_{\|}=k^2_{\|}+k'^2_{\|}-2k_{\|}k'_{\|}\cos\varphi,\quad
k'^2_{\|}+k'^2_{\perp}=k^2-2m\omega.
\eeq

For the phonon model one can easily proceed further using
delta-function in $\Lambda^{ph}$ and the small value of
$q^2=q^2_{\|}+q^2_{\perp}=\omega^2/s^2\ll k^2$. Integration gives
\beq{9.12}
\frac{dw_{ie}^{ph}}{d\omega}=
\frac{2}{\pi}\,
\frac{T}{Ms^2}\,
\frac{k_0\beta}{U}\,
\frac{v_{\perp}}{s}\,
\sqrt{\frac{\epsilon'-\epsilon_{\|}}{U}},
\eeq
where $k_0=\sqrt{2mU/\hbar^2}$ is the momentum at the potential
barrier $U$, $v_{\perp}=v\cos\theta$ is the normal component of
the incident neutron velocity, and
$\epsilon_{\|}=\epsilon\sin^2\theta$ is the energy related to the
incident neutron motion along the surface plane.

For the hydrodynamic model we can perform in (\ref{9.10})
integration over $q_{\perp}$ by closing integration path in
complex $q_{\perp}$ plane and suming over pole residues. Then it
is useful to write the result in the form similar to (\ref{9.12})
\beq{9.13}
\frac{dw_{ie}^{hyd}}{d\omega}=
\frac{2}{\pi}\,
\frac{\alpha T}{Ms^2}\,
\frac{k_0\beta}{U}\,
f(\epsilon,\theta,\epsilon',d),
\eeq
where dimensionless function $f$, depending on $\epsilon$,
$\theta$, $\epsilon'$ and parameter $d=2mD/\hbar$, is given by
\beq{9.14}
f(\epsilon,\theta,\epsilon',d)=
\frac{2}{\pi d}\,\frac{v_{\perp}}{v_0}\,
\int\limits_0^{\sqrt{k^2-2m\omega}}k'^2_{\perp}d k'_{\perp}
\int\limits_0^{\pi}d\varphi
L(k'_{\perp},\varphi).
\eeq
Here $v_0=\hbar k_0/m$ is the boundary neutron velocity, and
\beq{9.15}
\begin{array}{l}
L(k'_{\perp},\varphi)=
\Frac{\sqrt{2}}{\Gamma^3}\,
\Frac{(1-4\lambda^2)^{3/4}}{1-2\lambda^2}\times{}
\\[\bigskipamount]
\phantom{}\times
\left(\left(1+\Frac{1}{\mu}\right)
\Frac{\mu^2+\lambda^2}{1+2\mu+2(\mu^2-\lambda^2)}-
\Frac{\lambda(\lambda+1)}{1+2\lambda}\right),
\end{array}
\eeq
\beq{9.16}
2\lambda^2=\frac{q^2_{\|}}{q^2_{\|}+\sqrt{q^4_{\|}+\Gamma^4}},
\quad
2\mu^2=\frac{(\kp+\kp')^2}{q^2_{\|}+\sqrt{q^4_{\|}+\Gamma^4}}.
\eeq
To demonstrate the magnitude of the function $f$ (\ref{9.14}) and
its dependance on the final neutron energy $\epsilon'$ and
parameter $d$ we have performed numerical calculation of double
integral for typical values of initial neutron energy
$\epsilon=U/2$ and angle of incidence $\theta=\pi/4$. The results
are shown by solid lines in Fig.\ref{fig:1} for $d\ll 1$ and in
Fig.\ref{fig:2} for $d\ge 0.1$.

\begin{figure}
\vspace{10cm}
\caption{Function \protect{$f(\epsilon,\theta,\epsilon',d)$} from
(\ref{9.14}) for fixed initial neutron energy $\epsilon=U/2$ and
angle of incidence $\theta=\pi/4$ versus final neutron energy
$\epsilon'$ and parameter $d$.}
\label{fig:1}
\end{figure}

\begin{figure}
\vspace{10cm}
\caption{Function $f(\epsilon,\theta,\epsilon',d)$ for fixed
$\epsilon=U/2$ and $\theta=\pi/4$ versus final neutron energy
$\epsilon'$ and parameter $d$. Solid lines -- result of exact
numerical calculation (\ref{9.14}), dash lines -- approximation
(\ref{9.18}).}
\label{fig:2}
\end{figure}

It is seen that the spectrum of inelastically scattered neutrons
has a peak in the vicinity of the initial neutron energy. In this
region parameter
$\Gamma=\sqrt{|\omega|/D}=k_0\sqrt{|\epsilon'-\epsilon|/(Ud)}$ is
small with respect to $\kp+\kp'$ and the main contribution into
the integral comes from small
$q^2_{\|}(k'_{\perp},\varphi)$. Then one may simplify the function
$L$ (\ref{9.15}) by
taking limit $\mu^2\gg (\lambda^2,1)$:
\beq{9.17}
L\simeq\frac{1}{\sqrt{2}\Gamma^3}\,
\frac{(1-4\lambda^2)^{3/4}}{1+2\lambda},
\eeq
which allows to evaluate in (\ref{9.14}) all parameter dependence
and obtain
\beq{9.18}
f(\epsilon,\theta,\epsilon',d)\simeq
C\,\frac{v_{\perp}}{v_0}\,
\frac{1}{\sqrt{d}}\,
\sqrt{\frac{\epsilon'-\epsilon_{\|}}{|\epsilon'-\epsilon|}}.
\eeq
Here $C=0.47$ is the value of a dimensionless integral.

Approximation (\ref{9.18}) is valid for
$|\epsilon'-\epsilon|=|\omega|\ll Ud$, and for large parameter
$d\gg 1$ (\ref{9.18}) may give a good estimate not only for the
small $\omega$ peak but for the whole subbarrier area (as seen
from Fig.\ref{fig:2}).

Dependence on $\omega$ is governed mostly by the parameter $d$.
For $d\ll 1$ the peak is more pronounced and, when $d$ decreases,
becomes more narrow and high but with fixed (independent on $d$)
area. Indeed, using approximation (\ref{9.18}) one obtains
\beq{9.19}
\begin{array}{l}
w_{ie}^{hyd}=
{\ds\int}\Frac{dw_{ie}^{hyd}}{d\omega}d\omega\simeq
\Frac{4C}{\pi}\,
\Frac{\alpha T}{Ms^2}\,
\Frac{k_{\perp}\beta}{U}\,
\Frac{1}{\sqrt{d}}\,
{\ds\int\limits_0^{Ud}}
\sqrt{\Frac{\epsilon_{\perp}}{\omega}}d\omega={}
\\[\bigskipamount]
\phantom{w_{ie}^{hyd}=
{\ds\int}\Frac{dw_{ie}^{hyd}}{d\omega}d\omega}=
\Frac{8C}{\pi}\,
\Frac{\alpha T}{Ms^2}\,
k_{\perp}\beta\,
\sqrt{\Frac{\epsilon_{\perp}}{U}}.
\end{array}
\eeq
Contribution to the peak from phonon model can be neglected since
it has smooth behaviour and contains a suppression factor
$v/s\sim 10^{-3}$.

For $d>1$ the small $\omega$ peak becomes less pronounced (see
Fig.\ref{fig:2}) and the probability for neutron to remain under
the barrier after inelastic scattering diminishes as
$1/\sqrt{d}$ (since in (\ref{9.19}) the upper limit of the
integral is now $U$).

\section{Conclusion}

The general theory of neutron scattering is presented, valid for
the whole domain of slow neutrons from thermal to ultracold. For
thermal and cold neutrons, when the multiple scattering in the
target can be neglected, the cross section is reduced to that
known for thermal neutrons, which is determined mostly by
correlation function for the target matter (section \ref{s6}).

For UCN the rescattering is the dominant process, but the theory
can be simplified by exploiting small parameter $\vkappa{\bf u}$,
i.e. the ratio of the amplitude of thermal vibrations (for solid
targets) or relaxation lengths (for liquids) to neutron wave
length. In zero order approximation in $\vkappa{\bf u}$ (that is
equivalent to the scattering on a target with infinitely heavy
unmovable nuclei) it follows the known equation for elastic
scattering of UCN (\ref{1.9}). Dynamical processes in the target
are taken into account in the next orders in $\vkappa{\bf u}$ and
result in inelastic scattering.

A detail analysis of inelastic scattering needs separate
publication. Here in section \ref{s9} a specific example was
considered: scattering with small energy transfer when scattered
neutron remains below potential barrier. This quantitative example
allows to make some conclusions.

The value of cross section is very sensitive to correlation
function used. Phonon model which gives main contribution for UCN
excitation into thermal region is quite ineffective for small
energy transfer when space-time correlation processes are
determined mostly by relaxation ("hydrodynamic") processes. So,
the first condition to obtain reasonable theoretical result for
cross section with small energy transfer is the choice of an
adequate correlation function.

The second factor that needs a reasonable physical modeling is
elastic potential suited for target matter in each specific
experiment. Consistence, uniformity, possible existence of surface
layers, presence of hydrogen and its distribution -- all that may
require a change of model potential and therefore the wave
functions in input and output channels that enter the cross
section.

All effects connected with neutron spin are outside of the scope
of this work. For target nuclei with non zero spin the scattering
length depends on spin-spin orientation. This would require only
replacement in all formulae of scattering length $\beta$ by
weighted average value. (The same prescription is valid for
isotope non-uniform target). For large wave length of UCN the
averaging is well justified.

Neutron spin interaction with target electrons ("magnetic
scattering") does not present any specific difficulty, but require
inclusion of new "spin-spin" correlation function.

Quite different feature have spin-flip processes. Physically
interesting is calculation of depolarization probability for
stored polarized UCN. This effect (as well as neutron capture
(\ref{5b.7})) belongs to incoherent processes. They are not
considered in this work. It is evident that incoherent processes
can be considered by simple perturbation theory with "elastic"
functions as zero approximation.
\bigskip\bigskip

{\large\bf Acknowledgements}
\bigskip

We are grateful to V.I.Morozov for valuable discussions. The work
was supported by RFBR grant 96-15-96548.
\bigskip\bigskip

{\large\bf Appendix}: {\large\bf An alternative expression for
scattering amplitude}
\bigskip

We start with transformation of equation (\ref{5.28}). First,
using the definition (\ref{5.26}) for the matrix
$G_{jj'}(\nu\nu')$ and the expression (\ref{5.19e}) for the matrix
$\zeta_{jj'}(\nu\nu')$ one obtains for $G_{jj'}(\nu\nu')$
$$
\begin{array}{l}
G_{jj'}(\nu\nu')=-\Frac{2\pi}{m}{\ds\sum_{{\bf q}}}
e^{i{\bf q}(\vrho_{\nu}-\vrho_{\nu'})}\times{}
\\[\medskipamount]
\phantom{}\times
\av{j|e^{-i({\bf k}-{\bf q}){\bf u}_{\nu}}
\hat D_q^{-1}
e^{i({\bf k}-{\bf q}){\bf u}_{\nu'}}|j'}.
\end{array} \eqno (A1)
$$
Then it is convenient to introduce the operators
$$
\begin{array}{l}
\overrightarrow{P^{\mathstrut}}_{jj'}(\nu)=
\av{j|e^{-i({\bf k}+i\nabla_{\nu}){\bf u}_{\nu}}|j'},
\\[\medskipamount]
\overleftarrow{P^{\mathstrut}}_{jj'}(\nu)=
\av{j|e^{i({\bf k}-i\nabla_{\nu}){\bf u}_{\nu}}|j'},
\end{array} \eqno (A2)
$$
where arrows on operators $P_{jj'}(\nu)$ denote direction for
gradients to act on functions of $\vrho_{\nu}$. Thus, after
evident transformation with the help of (\ref{5.30}) the matrix
$G_{jj'}(\nu\nu')$ takes the form
$$
G_{jj'}(\nu\nu')=
\sum_{j''}\overrightarrow{P^{\mathstrut}}_{jj''}(\nu)
G_{j''}(\nu\nu')\overleftarrow{P^{\mathstrut}}_{j''j'}(\nu').
\eqno (A3)
$$
Equation (\ref{5.28}) with the use of (A3) can be written as
$$
\begin{array}{l}
\psi_j(\nu)+{\ds\sum_{j'j''\nu'}}
\overrightarrow{P^{\mathstrut}}_{jj''}(\nu)G_{j''}(\nu\nu')
\overleftarrow{P^{\mathstrut}}_{j''j'}(\nu')\psi_{j'}(\nu')
\beta_{\nu'}={}
\\[\medskipamount]
\phantom{\psi_j(\nu)+\sum_{j'j''\nu'}
\overrightarrow{P^{\mathstrut}}_{jj''}(\nu)G_{j''}(\nu\nu')}=
\delta_{ij}e^{i{\bf k}\vrho_{\nu}}.
\end{array} \eqno (A4)
$$

Note the action of operators $P_{jj'}(\nu)$ on exponents with
${\bf k}$ and ${\bf k}'$
$$
\begin{array}{c}
\overrightarrow{P^{\mathstrut}}_{jj'}(\nu)e^{i{\bf k}\vrho_{\nu}}=
\delta_{jj'}e^{i{\bf k}\vrho_{\nu}},
\\[\medskipamount]
e^{-i{\bf
k}'\vrho_{\nu}}\overleftarrow{P^{\mathstrut}}_{jj'}(\nu)=
e^{-i{\bf k}'\vrho_{\nu}}
\av{j|e^{i({\bf k}-{\bf k}'){\bf u}_{\nu}}|j'}.
\end{array} \eqno (A5)
$$
The scattering amplitude (\ref{5.32}), due to (A5), can be
represented in the form
$$
f_{ij}({\bf k},{\bf k}')=
-\sum_{j'\nu}e^{-i{\bf k}'\vrho_{\nu}}
\overleftarrow{P^{\mathstrut}}_{jj'}(\nu)\psi_{j'}(\nu)\beta_{\nu}
. \eqno (A6)
$$

Now let act on all terms of (A4) by the operator
$\overrightarrow{P^{\mathstrut}}{}^{-1}_{j'j}(\nu)$ and sum over
$j$. The right-hand side, due to (A5), remains unchanged and we
obtain a new form of general equation (\ref{5.28}) where the
matrix $G_j(\nu\nu')$ is "open" from the left
$$
\begin{array}{l}
{\ds\sum_{j'}}
\overrightarrow{P^{\mathstrut}}{}^{-1}_{jj'}(\nu)\psi_{j'}(\nu)+
{\ds\sum_{j'\nu'}}G_j(\nu\nu')
\overleftarrow{P^{\mathstrut}}_{jj'}(\nu')\psi_{j'}(\nu')
\beta_{\nu'}={}
\\[\medskipamount]
\phantom{{\ds\sum_{j'}}
\overrightarrow{P^{\mathstrut}}{}^{-1}_{jj'}(\nu)\psi_{j'}(\nu)+
{\ds\sum_{j'\nu'}}G_j(\nu\nu')}=
\delta_{ij}e^{i{\bf k}\vrho_{\nu}}.
\end{array} \eqno (A7)
$$
We may now multiply (A7) from the left by
$\beta_{\nu}\bar\psi^{(0)}_j(\nu)$, which is a solution of
equation (\ref{7.19}), i.e.
$$
\sum_{\nu}\beta_{\nu}\bar\psi^{(0)}_j(\nu)G_j(\nu\nu')=
e^{-i{\bf k}'\vrho_{\nu'}}-\bar\psi^{(0)}_j(\nu'). \eqno (A8)
$$
Then summing over $\nu$ we get with the help of (A8)
$$
\begin{array}{l}
{\ds\sum_{j'\nu}}\beta_{\nu}\bar\psi^{(0)}_j(\nu)
\overrightarrow{P^{\mathstrut}}{}^{-1}_{jj'}(\nu)\psi_{j'}(\nu)+{}
\\[\medskipamount]
\phantom{}+
{\ds\sum_{j'\nu}}\left(e^{-i{\bf k}'\vrho_{\nu}}-
\bar\psi^{(0)}_j(\nu)\right)
\overleftarrow{P^{\mathstrut}}_{jj'}(\nu)\psi_{j'}(\nu)
\beta_{\nu}={}
\\[\medskipamount]
\phantom{{\ds\sum_{j'\nu}}e^{-i{\bf k}'\vrho_{\nu}}-
\bar\psi^{(0)}_j(\nu)}=
\delta_{ij}{\ds\sum_{\nu}}\beta_{\nu}\bar\psi^{(0)}_j(\nu)
e^{i{\bf k}\vrho_{\nu}}.
\end{array} \eqno (A9)
$$
From (A6) and (A9) it follows for the scattering amplitude
$$
\begin{array}{l}
f_{ij}({\bf k},{\bf k}')=
-\delta_{ij}{\ds\sum_{\nu}}\beta_{\nu}e^{i{\bf k}\vrho_{\nu}}
\bar\psi^{(0)}_i(\nu)-{}
\\[\medskipamount]
\phantom{f_{ij}({\bf k},{\bf k}')}-
{\ds\sum_{j'\nu}}\Bigl(\bar\psi^{(0)}_j(\nu)
\overleftarrow{P^{\mathstrut}}_{jj'}(\nu)\psi_{j'}(\nu)
\beta_{\nu}-{}
\\[\medskipamount]
\phantom{f_{ij}
({\bf k},{\bf k}'){\ds\sum_{j'\nu}}{\ds\sum_{j'\nu}}}-
\beta_{\nu}\bar\psi^{(0)}_j(\nu)
\overrightarrow{P^{\mathstrut}}{}^{-1}_{jj'}(\nu)
\psi_{j'}(\nu)\Bigr),
\end{array} \eqno (A10)
$$
where zero order term is explicitly extracted.

As the last step, we perform the action of the operators
$P_{jj'}(\nu)$ on $\bar\psi^{(0)}_j(\nu)$ and $\psi_j(\nu)$ and
arrive at desired relation between scattering amplitude and exact
solution $\psi_j(\nu)$ of the general equation (\ref{5.28})
$$
\begin{array}{l}
f_{ij}({\bf k},{\bf k}')=
-\delta_{ij}{\ds\sum_{\nu}}\beta_{\nu}e^{i{\bf k}\vrho_{\nu}}
\bar\psi^{(0)}_i(\nu)-{}
\\[\medskipamount]
\phantom{f_{ij}}-
{\ds\sum_{j'\nu}}\av{j|e^{i{\bf k}{\bf u}_{\nu}}
\Bigl(\bar\psi^{(0)}_j(\vrho_{\nu}+{\bf u}_{\nu})
\psi_{j'}(\vrho_{\nu})-{}
\\[\medskipamount]
\phantom{f_{ij}{\ds\sum_{j'\nu}}
{\ds\sum_{j'\nu}}{\ds\sum_{j'\nu}}}-
\bar\psi^{(0)}_j(\vrho_{\nu})
\psi_{j'}(\vrho_{\nu}-{\bf u}_{\nu})\Bigr)\beta_{\nu}|j'}.
\end{array}\eqno (A11)
$$
An expansion of (A11) in ${\bf u}_{\nu}$ gives in zero order (with
the use of (\ref{7.18}) and (\ref{7.10}))
$$
f_{ij}^{(0)}({\bf k},{\bf k}')=
-\delta_{ij}\psi(-{\bf k}',-{\bf k}), \eqno (A12)
$$
what, due to time reversal invariance, equals to (\ref{7.9}). The
first order term in (A11) coincides with (\ref{7.22}).

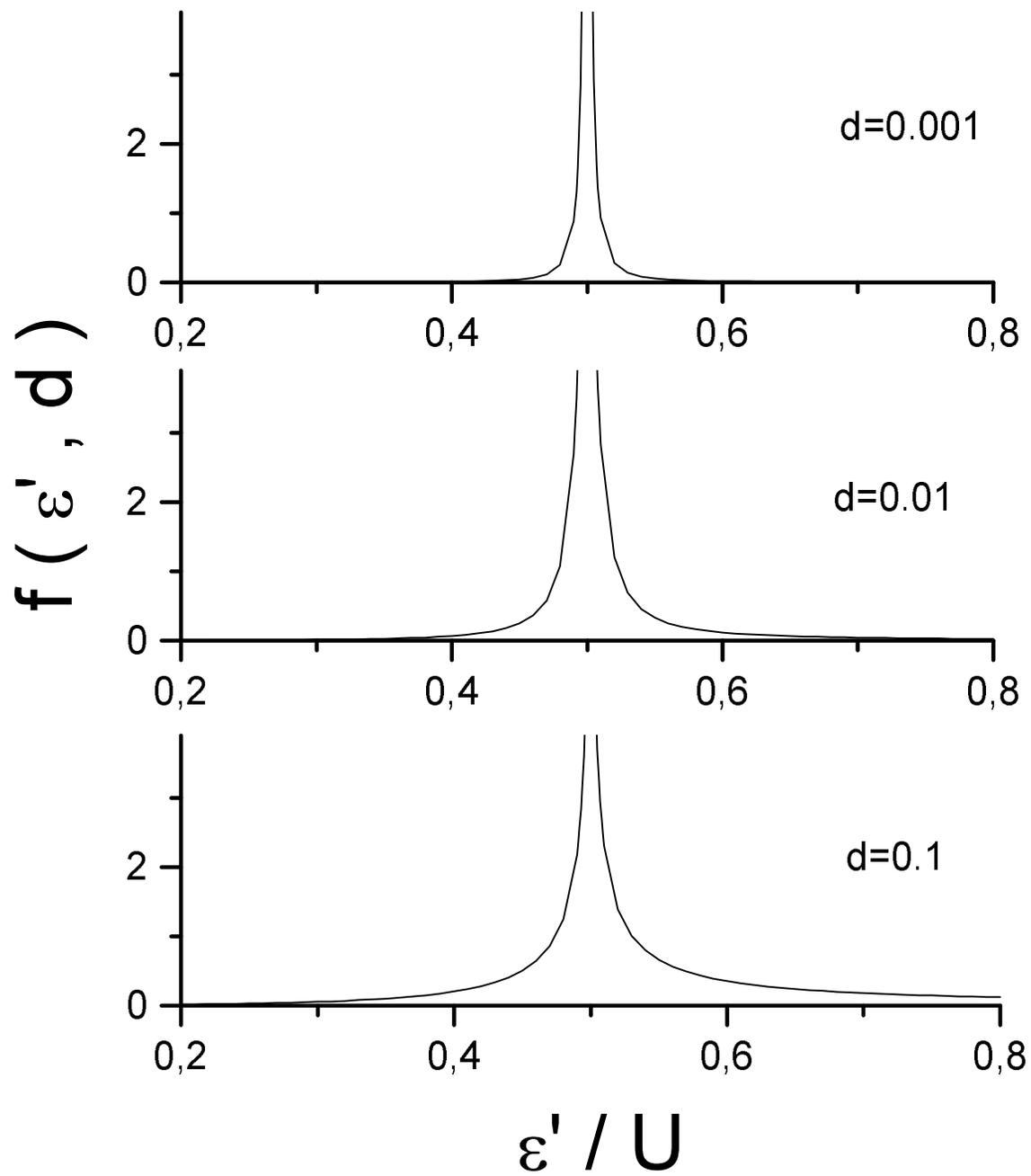
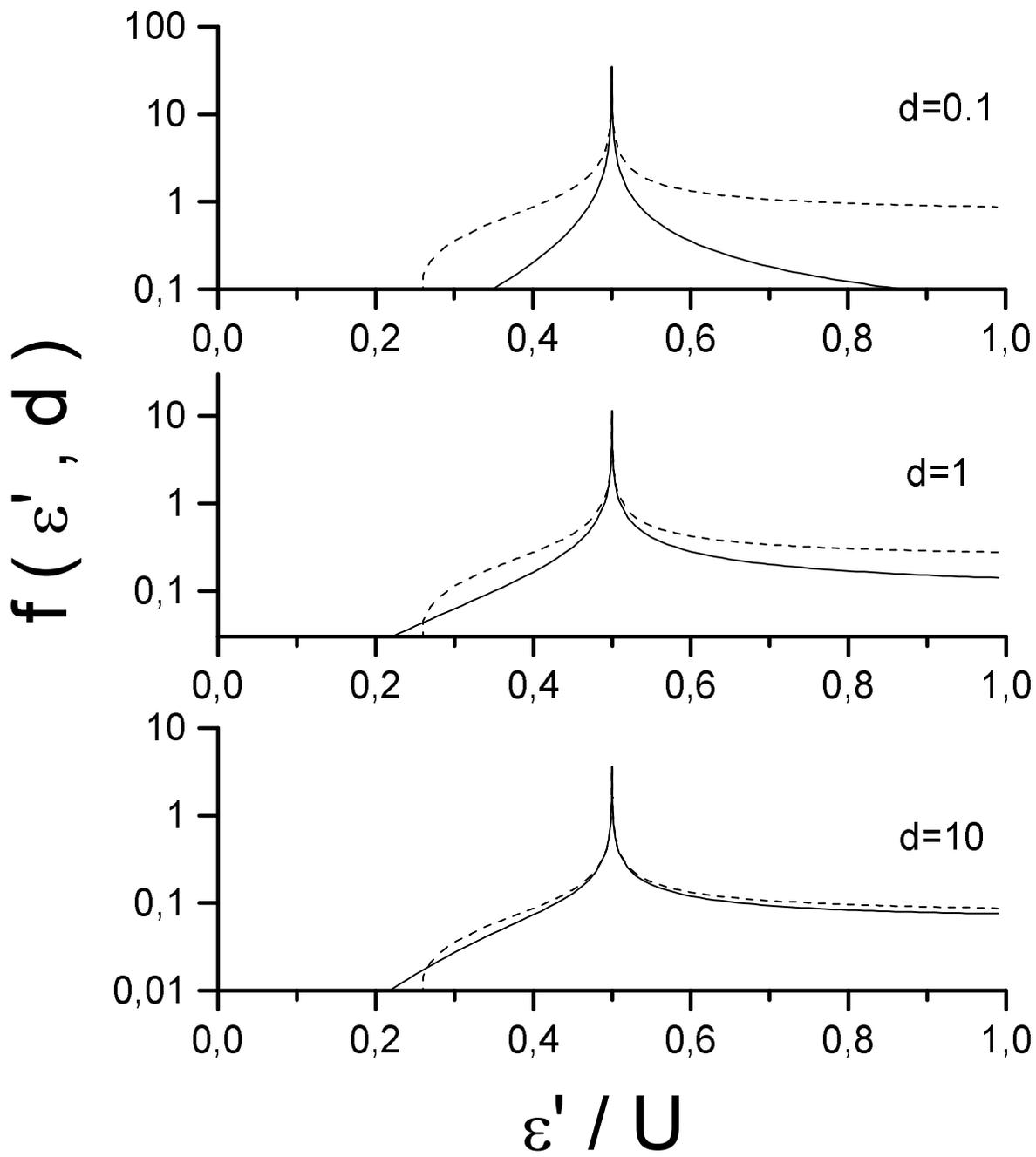
\end{document}